\documentclass[12pt]{article}
\pdfoutput=1
\usepackage[utf8]{inputenc}
\usepackage[T1]{fontenc}
\usepackage{float}
\usepackage{authblk}
\usepackage[normalem]{ulem}
\usepackage{xcolor}
\usepackage{amsmath}
\usepackage{amsthm}
\usepackage{amssymb}
\usepackage{amsfonts}
\usepackage{adjustbox}
\usepackage{bbm}
\usepackage{graphicx}
\usepackage{booktabs}
\usepackage{mathtools}
\usepackage{multirow,array}
\usepackage{fullpage}
\usepackage{comment}
\usepackage[mathlines]{lineno}
\usepackage{multirow}
\usepackage{subcaption}
\usepackage[style=numeric-comp,sorting=none,maxnames=99]{biblatex}

\usepackage[colorlinks = true,
            linkcolor = teal,
            urlcolor  = teal,
            citecolor = teal]{hyperref}

\title{Collective behavior from surprise minimization}
 
\author[a,b,c,e]{Conor Heins\thanks{\href{mailto:cheins@ab.mpg.de}{cheins@ab.mpg.de}}}
\author[d]{Beren Millidge}
\author[e,f,g]{Lancelot Da Costa}
\author[h]{Richard~P.~Mann}
\author[e,g]{Karl Friston} 
\author[a,b,c]{Iain D. Couzin\thanks{\href{mailto:icouzin@ab.mpg.de}{icouzin@ab.mpg.de}}} 

\affil[a]{Department of Collective Behaviour, Max Planck Institute of Animal Behavior, Konstanz D-78457, Germany}
\affil[b]{Centre for the Advanced Study of Collective Behaviour, University of Konstanz, Konstanz D-78457, Germany}
\affil[c]{Department of Biology, University of Konstanz, Konstanz D-78457, Germany}
\affil[d]{Medical Research Council Brain Networks Dynamics Unit, University of Oxford, Oxford OX1 3TH, UK}
\affil[e]{VERSES Research Lab, Los Angeles, CA 90016, USA}
\affil[f]{Department of Mathematics, Imperial College London, London SW7 2AZ, UK}
\affil[g]{Wellcome Centre for Human Neuroimaging, University College London, London, WC1N 3AR, UK}
\affil[h]{Department of Statistics, School of Mathematics, University of Leeds, Leeds, LS2 9JT, UK}

\date{}

\begin{document}
\maketitle
\pagenumbering{arabic}

\begin{abstract}
Collective motion is ubiquitous in nature; groups of animals, such as fish, birds, and ungulates appear to move as a whole, exhibiting a rich behavioral repertoire that ranges from directed movement to milling to disordered swarming. Typically, such macroscopic patterns arise from decentralized, local interactions among constituent components (e.g., individual fish in a school). Preeminent models of this process describe individuals as self-propelled particles, subject to self-generated motion and `social forces' such as short-range repulsion and long-range attraction or alignment. However, organisms are not particles; they are probabilistic decision-makers. Here, we introduce an approach to modelling collective behavior based on active inference. This cognitive framework casts behavior as the consequence of a single imperative: to minimize surprise. We demonstrate that many empirically-observed collective phenomena, including cohesion, milling and directed motion, emerge naturally when considering behavior as driven by active Bayesian inference --- without explicitly building behavioral rules or goals into individual agents. Furthermore, we show that active inference can recover and generalize the classical notion of social forces as agents attempt to suppress prediction errors that conflict with their expectations. By exploring the parameter space of the belief-based model, we reveal non-trivial relationships between the individual beliefs and group properties like polarization and the tendency to visit different collective states. We also explore how individual beliefs about uncertainty determine collective decision-making accuracy. Finally, we show how agents can update their generative model over time, resulting in groups that are collectively more sensitive to external fluctuations and encode information more robustly. 
\end{abstract}

\section*{Significance Statement}
We introduce a model of collective behavior, proposing that individual members within a group, such as a school of fish or a flock of birds, act to minimize surprise. This active inference approach naturally generates well-known collective phenomena such as cohesion and directed movement without explicit behavioral rules. Our model reveals intricate relationships between individual beliefs and group properties, demonstrating that beliefs about uncertainty can shape collective decision-making accuracy. As agents update their generative model in real-time, groups become more sensitive to external perturbations and more robust in encoding information. Our work provides fresh insights into understanding collective dynamics and could inspire strategies in the study of animal behavior, swarm robotics and distributed systems.

\section*{Introduction}

The principles underlying coordinated group behaviors in animals have inspired research in disciplines ranging from zoology to engineering to physics \cite{major1978three, camazine2003self, rubenstein2012kilobot}. Collective motion in particular has been a popular phenomenon to study, due in part to its striking visual manifestation and ubiquity (e.g., swarming locusts, schooling fish, flocking birds and herding ungulates), and in part to the simplicity of models that can reproduce many of its qualitative features; like cohesive, directed  movement \cite{Aoki1982,reynolds1987flocks,vicsek1995novel, couzin2002collective}. Because of this, collective motion is often cited as a canonical example of a self-organizing complex system, wherein collective properties emerge from simple interactions among distributed components.

Popular theoretical models cast collective motion as groups composed of self-propelled particles (SPPs) that influence one another via simple `social forces.' Early models like the Vicsek model \cite{vicsek1995novel} consider only a simple alignment interaction, where each particle aligns its direction of travel with the average heading of its neighbors. While oversimplifying the biological mechanisms in play, SPP models --- like the Vicsek model --- are useful for their amenability to formal understanding, e.g., the computation of universal quantities and relations through hydrodynamic and mean-field limits \cite{toner1998flocks,sumpter2006principles, bertin2006boltzmann, degond2008continuum}.

Recent research has shifted towards more biologically-motivated approaches that aim to model the specific behavioral circuits and decision-rules that govern individual behaviors \cite{herbert2011inferring,calovi2014swarming,hein2020algorithmic,fahimipour2023wild}. While these models are less analytically-tractable than SPP models, they are more appealing to domain specialists like biologists, as they can generate predictions about sensory features in an individual's environment that are necessary and sufficient for evoking behavior. Furthermore, these predictions can be tested experimentally \cite{gautrais2012deciphering, hein2020algorithmic}. This data-driven approach can thus provide mechanistic insights into the biological and cognitive origins of decision-making \cite{katz2011inferring,calovi2014swarming}.

In this work, we propose a model class that blends the first-principles, theoretical approach of physical models with biological-plausibility, resulting in an ecologically-valid but theoretically-grounded agent-based model of collective behavior. Our model class is based on active inference, a framework for designing and describing adaptive systems where all aspects of cognition --- learning, planning, perception, and action --- are viewed as a process of inference \cite{friston2009reinforcement,friston2017active,parr2022active,da2020active}. Active inference originated in theoretical neuroscience as a normative account of self-organizing, biological systems as constantly engaged in predictive exchanges with their sensory environments \cite{friston2005theory, friston2006free, friston2011optimal,fristonFreeEnergyPrinciple2023a}.
\section*{Collective motion models: from self-propelled particles to Bayesian agents}

In popular self-propelled particle models, an individual's movement is described as driven by a combination of social and environmental forces. These forces are often treated as vectors that capture various tendencies seen in biological collective motion, such as repulsion, attraction (to neighbors or external targets), and alignment. These forces can then be combined with various nonlinearities and weights to capture mechanisms of interaction. 

\begin{figure*}
\centering
\includegraphics[width=0.8\linewidth]{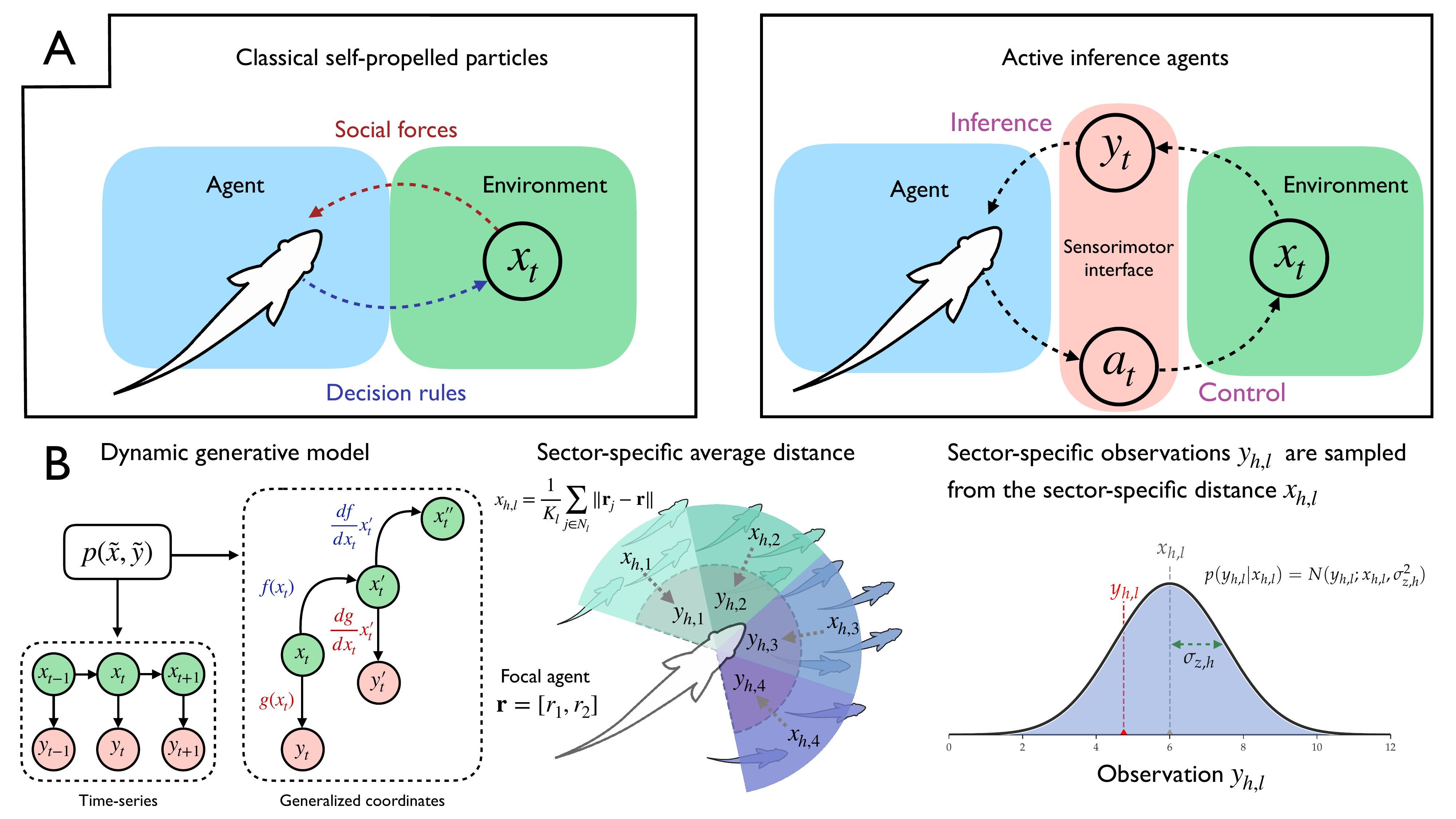}
\caption{\textbf{A}: Schematic illustrating the Bayesian perspective in the context of our single agents, where the hidden states of the environment are segregated from a focal agent by means of sensory data $y_t$ (right panel of \textbf{A}). This contrasts with classic self-propelled particle models (left panel of \textbf{A}), where environmental or social information manifests in terms of social forces on the focal individual, who emits its own actions based on hand-crafted decision-rules (e.g., changes to heading direction). \textbf{B}: Schematic illustration of the sector-specific distance tracking. The left panel shows a Bayesian network representation of a dynamic generative model (i.e., a time-series model), that represents the time-evolution of a latent variable $x_{1,...,T}$ and simultaneous observations $y_{1,...T}$. Shown are both a standard time-series representation (lower left) and its equivalent representation as generalized coordinates of motion $\tilde{x}_t = (x_t, x_t', x_t'', ... )$ (right). We show the orders of differentiation used for our model in practice (3 orders of motion for $\tilde{x}$ and 2 orders of motion for $\tilde{y}$). The middle panel of \textbf{B} shows how each component of the vectorial hidden state $\mathbf{x} = (x_{h,1}, ..., x_{h,L})$ is computed as the average nearest-neighbor distance for the neighbors within each visual sector. Observations are generated as noisy, Gaussian samples centered on the sector-wise distance hidden state (right panel of \textbf{B}). This requires the agent to estimate the true hidden state $x_t$ by performing inference with respect to a generative model of how sensory data are generated $p(\tilde{\mathbf{y}}, \tilde{\mathbf{x}})$.}
\label{fig:AIF_vs_SPP_scheme_fish_perception_schematic}
\end{figure*}

In contrast, the active inference approach forgoes specifying explicit vectorial forces, and instead starts by modelling all behavior as the solution to an inference problem, namely the problem of inferring the latent causes of sensations. Perception and action strive to improve the agent's predictions of sensory inputs, based on its internal model of its world (see Figure \ref{fig:AIF_vs_SPP_scheme_fish_perception_schematic}A). By equipping this internal model with expectations about the environment's underlying tendencies, `social forces' can emerge naturally as agents attempt to suppress sensory data that are mismatched with their expectations. This perspective-shift offers a unifying modelling ontology for describing adaptive behavior, while also resonating with cybernetic principles like homeostatic regulation and process theories of neural function like predictive coding \cite{rao1999predictive, adams2013predictions,ali2022predictive, caucheteux2023evidence}.

Active inference blends the construct validity of cognitivist approaches with the first-principles elegance of physics-based approaches by invoking minimization of a single, all-encompassing objective function that explains behavior: surprise, or, under certain assumptions, prediction error. As an example of this perspective shift, in this work we investigate a specific class of generative models that can be used to account for the types of collective behaviors exhibited by animal groups. In doing so, we hope to showcase the benefits of the framework, while also proposing a testable model class for use in studies of biological collective motion.

\section*{Active inference and generative models of behavior}

A common pipeline in the quantitative study of animal behavior involves selecting a candidate behavioral algorithm or decision rule that may explain a given behavior, and then fitting the parameters of the candidate model to experimental or observational data \cite{laland2004social, gautrais2012deciphering}. While these approaches often yield strong quantitative fits to data, the explanatory power of the models reduces to the interpretation of hard-coded parameters, which often have opaque relationships to real biological mechanisms or constructs \cite{krafft2021bayesian}. 

In the active inference framework we rather ask: what is the minimal model an organism might have of its environment that is sufficient to explain its behavior? Behavior is  cast as the process by which the agent minimizes surprise or prediction error, with respect to this model of the world \cite{friston2005theory,baltieri2019generative}. The principle of prediction-error minimization enjoys empirical support in neuroscience \cite{rao1999predictive, URAN20221240} and a theoretical basis in the form of the Free Energy Principle \cite{friston2005theory, friston2006free,fristonFreeEnergyPrinciple2023a}, an account of all self-organizing systems that casts them as implicit models of their environments, ultimately in the service of minimizing the surprise (a.k.a., self-information) associated with sensory states \cite{friston2009free,hohwy2016self,friston2019free}.

What states-of-affairs count as surprising hinges on a generative model that can assign a likelihood to sensory data. When it comes to modelling behavior driven by this principle, the challenge then becomes specifying a generative or world model, whereby a particular pattern of behavior simply emerges by minimizing surprise.

According to active inference, agents minimize surprise by changing their beliefs about the world (changing which observations are considered surprising) or by acting on the world to avoid surprising sensory data. The former strategy is thought to correspond to passive processes such as perception and learning, whereas the latter corresponds to processes like active sensing and movement. Action is thus motivated by the desire to generate sensations that are as least surprising as possible.

In this paper, we describe the motion of mobile, mutually-sensing agents as emerging from a process of collective active inference, whereby agents both estimate the hidden causes of their sensations, while also actively changing their position in space in order to minimize prediction error. In contrast to models that use pre-specified behavioral rules for generating behavior, generative models entail collective behavior by appealing to a probabilistic representation of how an organism's sensory inputs are generated.

\section*{A generative model for a (social) particle}

We now consider a sufficient generative model for an individual in a moving group. We equip this individual, hereafter referred as the focal agent, with a representation of a simple random variable: the local distance $x$ between itself and its neighbors. For generality, we can expand this into a multivariate random variable to describe a set of distances $\mathbf{x} = (x_1, x_2, ..., x_L)$ that track the distance between the focal agent and its neighbors within $L$ different sensory sectors (see Figure \ref{fig:AIF_vs_SPP_scheme_fish_perception_schematic}B). We analogize these $L$ sectors to adjacent visual fields of an agent's field of view \cite{collignon2016stochastic, bastien2020model}. 

The focal agent possesses a model of the distance(s) $\mathbf{x}$ and its sensations thereof $\mathbf{y}$. In particular, our focal agent represents the dynamics of $\mathbf{x}$ using a stochastic differential equation (a.k.a., a state-space model) defined by a drift $\mathbf{f}$ and some stochastic forcing $\omega$ --- we refer to this component of the generative model as the dynamics model. The stochastic term $\omega$ captures the agent's uncertainty about paths of $\mathbf{x}$ over time. The agent also believes it can sense $\mathbf{x}$ via observations $\mathbf{y}$, mediated by a sensory map, which we  call the observation model. This is defined by some (possibly non-linear) function $\mathbf{g}$ with additive noise $z$. The agent's generative model is then fully described by a pair of equations that detail 1) the time-evolution of the distance and 2) the simultaneous generation of sensory samples of the distance:

\begin{align}
    D \tilde{\mathbf{x}} &= \tilde{\mathbf{f}} + \tilde{\boldsymbol \omega} & \tilde{\mathbf{y}} &= \tilde{\mathbf{g}} + \tilde{\boldsymbol z} 
\end{align}

All random variables are described using generalized coordinates of motion with the convention $\tilde{\mathbf{q}} = 
\{\mathbf{q}, \mathbf{q}', \mathbf{q}'', ...\}$. Generalized coordinates allow us to represent the trajectory of a random variable using a vector of local time derivatives (position, velocity, acceleration, etc.). The matrix $D$ is a generalized derivative operator that moves a vector of generalized coordinates up one order of motion $D(x, x', x'', ...)^{\top} = (x', x'', x''', ...)^{\top}$. The generalized functions $\tilde{\mathbf{f}}$ and $\tilde{\mathbf{g}}$ therefore operate on vectors of generalized coordinates (see \textit{SI Appendix}, Section A for details on generalized coordinates and filtering).

\section*{Generalized filtering and active inference}

An agent equipped with this dynamic generative model then performs active inference by updating its beliefs (state estimation, or filtering) and control states (action) to minimize surprise.

Inference entails updating a probabilistic belief over hidden states $\tilde{\mathbf{x}}$ in the face of sensory data $\tilde{\mathbf{y}}$. Our agents solve this filtering problem using generalized filtering \cite{friston2010generalised,friston2007variational}, an algorithm for approximate Bayesian inference and parameter estimation on dynamic state-space models. This is achieved by minimizing the variational free energy $F$, a tractable upper bound on surprise (i.e., negative log evidence or marginal likelihood). The agent minimizes the free energy with respect to a belief distribution $q(\tilde{\mathbf{x}})$ with parameters $\nu$; this approximates the true posterior $q_{\nu}(\tilde{\mathbf{x}}) \approx p(\tilde{\mathbf{x}} | \tilde{\mathbf{y}})$, which is the optimal solution in the context of Bayesian inference. The true posterior $p(\tilde{\mathbf{x}} | \tilde{\mathbf{y}})$ is difficult to compute for many generative models due to the difficult calculation of the marginal (log) likelihood $\ln p(\tilde{\mathbf{y}})$. Variational methods circumvent this intractable marginalization problem by replacing it with a tractable optimization problem: namely, adjusting an approximate posterior to match the true posterior by minimizing $F$ with respect to its (variational) parameters $\nu$.

We parameterize $q(\tilde{\mathbf{x}})$ as a Gaussian with mean-vector $\tilde{\boldsymbol \mu}$, which is a natural choice for this generative model since the assumption of normally-distributed noises $\tilde{\mathbf{z}}, \tilde{\boldsymbol \omega}$ imply that the true posterior will be Gaussian near the posterior mode $\arg \max p(\tilde{\mathbf{x}}|\tilde{\mathbf{y}})$. The implicit Gaussian (i.e., Laplace) assumption is ubiquitous in the modelling and signal processing literature \cite{mackay2003information} and can be regarded as a `minimal' assumption, by appeal to things like the central limit theorem and related principles (e.g., Jaynes' maximum entropy principle). According to generalized filtering, $\tilde{\boldsymbol \mu}$ is updated using a sum of prediction errors:

\begin{align}
    \frac{d \tilde{\boldsymbol{\mu}}}{d t} &\propto -\nabla_{\tilde{\boldsymbol{\mu}}} F(\tilde{\boldsymbol \mu}, \tilde{\mathbf{y}}) \notag \\\
    &\propto \tilde{\boldsymbol{\varepsilon}}_{z} - \tilde{\boldsymbol{\varepsilon}}_{\omega} \notag \\\
    \textrm{where} \hspace{2mm}  \tilde{\boldsymbol{\varepsilon}}_{z} &=\tilde{\mathbf{y}} - \tilde{\mathbf{g}}(\tilde{\boldsymbol \mu})\notag \\\
    \tilde{\boldsymbol{\varepsilon}}_{\omega} &=  D\tilde{\boldsymbol\mu} - \tilde{\mathbf{f}}(\boldsymbol{
    \tilde{\mu}})
    \label{eq:PC_update_dynamic_gm}
\end{align}

The ensuing evidence accumulation can be regarded as a natural generalization of predictive coding \cite{rao1999predictive,jehee2006learning, spratling2017review}, where beliefs about local trajectories $\tilde{\boldsymbol \mu}$ are updated using a running assimilation of sensory and model prediction errors: 
$\tilde{\boldsymbol{\varepsilon}}_z$ and  $\tilde{\boldsymbol{\varepsilon}}_{\omega}$, respectively. For notational clarity, we have omitted terms that weigh these prediction errors; the so-called generalized sensory and model precisions $\tilde{\Pi}^{\textbf{z}}, \tilde{\Pi}^{\boldsymbol \omega}$, which encode the agent's assumptions about the magnitude and correlation structure of noise. The importance of these precisions will become clear later, when understanding the relationship between precision-weighted prediction errors and social forces.

While inference entails changing the approximate posterior means $\tilde{\boldsymbol{\mu}}$ to best explain sensory data, action entails changing the data itself to better match the data to one's current beliefs. Similar to the update scheme in \eqref{eq:PC_update_dynamic_gm}, actions are also updated by minimizing free energy:

\begin{align}
    \frac{da}{dt} &= -\nabla_a F(\tilde{\boldsymbol \mu}, \tilde{\mathbf{y}}(a))\notag \\\
    &= -\nabla_{\tilde{\mathbf{y}}} F(\tilde{\boldsymbol \mu},\tilde{\mathbf{y}}(a) ) \nabla_{a} \tilde{\mathbf{y}}(a) \notag \\\
    &\propto -\tilde{\boldsymbol{\varepsilon}}_{z}^{\top}\nabla_a \tilde{\mathbf{y}}(a)  \label{eq:action_update} 
\end{align}

Actions thus are updated using a product of sensory prediction errors $\tilde{\boldsymbol{\varepsilon}}_{z}$ and a `sensorimotor contingency' $\nabla_{a} \tilde{\mathbf{y}}(a)$ or reflex arc. This sort of `reflexive action' --- where control is simply targeted at minimizing sensory prediction errors --- underlies active inference accounts of motor control \cite{adams2013predictions, maselli2022active}, and can be formally related to proportional-integral-derivative (PID) control \cite{baltieri2019pid}. These prediction errors measure how far an agent's observations are from its expectations; the agent then acts using \eqref{eq:action_update} to minimize this deviation. Active inference agents are thus driven to act in a way that aligns with their (biased) expectations about the world \cite{buckley2017free}. In the next section, we will see how building a particular type of bias into each agent's generative model leads to the appearance terms in \eqref{eq:action_update} that resemble social forces.

\section*{Social forces as a consequence of predictive control}
 
In particular, we take the agent's action to be its heading direction $a = \mathbf{v}$, and examine the case where the agent observes the distance to its neighbors within a single sensory sector, i.e., $L = 1$, $\mathbf{x} = (x_1)$. We distinguish the agent's representation of the distance $\mathbf{x}$ from the actual distance using the subscript $h$. Therefore $\mathbf{x}_h =(x_{h,1}, x_{h,2}, ..., x_{h,L})$ denotes the average distances (and corresponding sensory samples $\mathbf{y}_h$) calculated using the actual positions of other agents. For the case of $L=1$, and assuming the agent observes both the distance and its rate of change $y'_{h,1}$, this is:

\begin{align}
\label{eq:gen_process_x_and_y}
x_{h,1} &= \frac{1}{K}\sum_{j\in N_{in}} \| \mathbf{r}_j - \mathbf{r}\| &y_{h,1} &= x_{h,1} + z_{h,1} \notag \\
x'_{h,1} &= \frac{dx_{h,1}}{dt}  & y'_{h,1} &= x'_{h,1} + z'_{h,1}
\end{align}

$N_{in}$ is the set of neighbors within the agent's single sensory sector, $K$ is the size of this set,  $\mathbf{r}$ is the focal agent's position vector, and $\mathbf{r}_j$ is the position vectors of neighbor $j$. The sensory observation of the generalized distance $\tilde{y}_h = (y_{h,1}, y'_{h,1})$ is a sample of the hidden state, perturbed by some additive noises $\tilde{z} = (z_{h,1}, z'_{h,1})$. By expanding the active inference control rule in \eqref{eq:action_update}, we arrive at the following differential equation for the heading vector:

\begin{align}
    \frac{d\mathbf{v}}{dt} &= \xi'_{z} \Delta \hat{\mathbf{r}}
  \notag \\
    \xi'_{z} &= \pi'_{z,1} (y'_{h,1} - \mu'_{h,1}) \notag \\
    \Delta \hat{\mathbf{r}} &= \frac{1}{K}\sum_{j\in N_{in}}\frac{\Delta \mathbf{r}_{j}}{\| \Delta \mathbf{r}_{j} \|}, \Delta \mathbf{r}_{j} = \mathbf{r}_j - \mathbf{r} \label{eq:expanded_action_rule}
\end{align}

The average vector $\Delta \hat{\mathbf{r}}$ is exactly the (negative) `sensorimotor contingency' term $\nabla_{a}\tilde{\mathbf{y}}(a)$ from \eqref{eq:action_update} (see \textit{SI Appendix}, Section A for detailed derivations):

\begin{align}
    \nabla_{\mathbf{v}} \tilde{y}(\mathbf{v}) &= \nabla_{\mathbf{r}} y = \frac{1}{K}\sum_{j\in N_{in}}\frac{\mathbf{r} - \mathbf{r}_j}{\| \mathbf{r} - \mathbf{r}_j \|} = -\Delta \hat{\mathbf{r}} \label{eq:derivation_SMC}
\end{align} 

The simple action update in \eqref{eq:expanded_action_rule} means that the focal agent moves along a vector pointing towards the average position of its neighbors. Whether this movement is attractive or repulsive is determined by the sign of the precision-weighted prediction error $\xi'_{z}=\pi'_{z,1} (y'_{h,1} - \mu'_{h,1})$, and its magnitude depends on two factors: 1) the sensory precision or `reliability' $\pi'_{z,1}$ that the agent affords observations of the rate-of-change of $y_{h,1}$; and 2) the degree to which these rate-of-change observations deviate from their predicted value $y'_{h,1} - \mu'_{h,1}$.

The presence of both attractive and repulsive forces depends on the agent's model of the distance dynamics, captured by the functional form of $\tilde{\mathbf{f}}$. In particular, consider forms of $\tilde{\mathbf{f}}$ that relax $\mathbf{x}$ to some attracting fixed point $\eta > 0$. Equipped with such a stationary model of the local distance, inference dynamics (c.f., \eqref{eq:PC_update_dynamic_gm}) will constantly bias its predictions $\mu$ according to the prior belief that the distance is pulled to $\eta$. Given this biased dynamics model and the action update in \eqref{eq:action_update}, such an agent will move to ensure that distance observations $\tilde{y}_h$ are equal to the fixed point $\eta$.

This action update shows immediate resemblance to the attractive and repulsive vectors common to social force-based models \cite{couzin2002collective, Aoki1982, reynolds1987flocks}, which often share the following general form:

\begin{align}
    F_{attr} &\propto \sum_{j \in Z_{A}}\frac{\mathbf{r}_{ij}}{\| \mathbf{r}_{ij} \|} \notag \\
    F_{repul} &\propto -\frac{1}{K}\sum_{j \in Z_{R}}\frac{\mathbf{r}_{ij}}{\| \mathbf{r}_{ij} \|} \notag \\
\end{align}

where $Z_{A}, Z_{R}$ refer to distance-defined zones of attraction or repulsion, respectively. In the active inference framework, these social forces emerge as the derivative of the observations with respect to action $\nabla_{a} \tilde{y}$, where the sign and magnitude of the precision-weighted sensory prediction error $\xi'_{z}$ determines whether the vector is attractive (towards neighbors) or repulsive (away from neighbors). The transition point between attraction and repulsion is therefore given by $\eta$, the point at which prediction errors switch sign.

An important consequence of this formulation is that, unlike the action rule used in social force-based models, the `steady-state' solution occurs when all social forces disappear (when prediction errors vanish). In this case, the agent ceases to change its heading direction and adopts its previous velocity. This occurs when the agent's sensations align with its (biased) predictions $y_{h,1} \approx \eta$. In classic SPP models, this is equivalent to the different social force vectors exactly cancelling each other.

We can therefore interpret social force-based models as limiting cases of distance-inferring active inference agents, because one can conceive of social forces as just those forces induced by free energy gradients; namely, the forces that drive belief-updating. In the case of our active inference agents, attractive and repulsive forces emerge naturally when we assume A) agents model the local distance dynamics as an attractor with some positive-valued fixed point $\eta$; B) agents can act by changing their heading direction and C) agents observe at least the first time derivative of their observations (e.g., $y'_{h,1}$, but see \textit{SI Appendix}, Section A for detailed derivations).

It is worth highlighting the absence of an explicit, vectorial alignment force in this model, consistent with experimental findings in two species of fish \cite{herbert2011inferring,katz2011inferring}. The heading vectors of neighbors is nevertheless implicitly incorporated into the calculation of first-order prediction errors $\xi'_{z}$ via the first order hidden state $x'_{h,1}$ (c.f., \eqref{eq:gen_process_x_and_y} and \textit{SI Appendix}, Eq.\eqref{eq:generalised_hidden_states_GP}). In particular, the $x'_{h,1}$ (from which the observations $y'_{h,1}$ are sampled) is equivalent to the `relative velocity' term used in so-called selective attraction and repulsion models, where the instantaneous rate at which neighbours approach or move away, is used to drive movement \cite{romanczuk2009collective}. However, explicit alignment forces as seen in the Vicsek model \cite{vicsek1995novel} and 2- and 3-Zone Couzin models \cite{couzin2002collective, couzin2005effective} can also be recovered if we assume agents have a generative model of the average angle between their heading vector and those of their neighbors (see \textit{SI Appendix}, Section B for derivations of alignment forces).

\section*{Multivariate sensorimotor control}
Having recovered social forces as free energy gradients in the case of a single sensory sector ($L=1$), we now revisit the general formulation of the generative model's state-space, where the hidden variable $x$ is treated as an $L$-dimensional vector state: $\mathbf{x} = (x_1, x_2, ..., x_L)$, with correspondingly $L$-dimensional observations $\mathbf{y} = (y_1, y_2, ..., y_L)$. 

Specifically, we consider each $x_l$ to represent the average distance-to-neighbors within one of a subset of adjacent sensory sectors, where each sector is offset from the next by a fixed inter-sector angle (see Figure \ref{fig:AIF_vs_SPP_scheme_fish_perception_schematic}B for a schematic of the multi-sector set-up). The rest of the generative model is identical; the agents estimate these distances (and their temporal derivatives $x_l', x_l'', ...$) while changing their heading direction to minimize free energy. Following the same steps as in the case of a single sector, the resulting update rule for $\mathbf{v}$ is a weighted sum of `sector-vectors', where generalized observations from each sector-specific modality $\tilde{y}_l$ are used to compute the prediction errors that scale the corresponding sector-vector. This generalizes the scalar-vector product in \eqref{eq:expanded_action_rule} to a matrix-vector product:

\begin{align}
    \frac{d\mathbf{v}}{dt} &= \tilde{\boldsymbol \xi}_{z}^{\top} \Delta \hat{\mathbf{R}}  \notag \\
    \Delta \hat{\mathbf{R}} &= -\begin{bmatrix} \nabla_{\mathbf{v}} \tilde{y}_1 \\ \nabla_{\mathbf{v}} \tilde{y}_2 \\ \vdots \\ \nabla_{\mathbf{v}} \tilde{y}_L \end{bmatrix}\label{eq:multivariate_action}
\end{align}

where now the (negative) sensorimotor contingency $-\nabla_{a} \tilde{\mathbf{y}} = \Delta \hat{\mathbf{R}}$ is a matrix whose rows contain the partial derivatives $\nabla_{\mathbf{v}} \tilde{y}_l$ (i.e. the `sector-vectors'). Each sector-vector is a vector pointing towards the average neighbor position within sector $l$. 

\begin{figure*}
\centering
\includegraphics[width=0.85\linewidth]{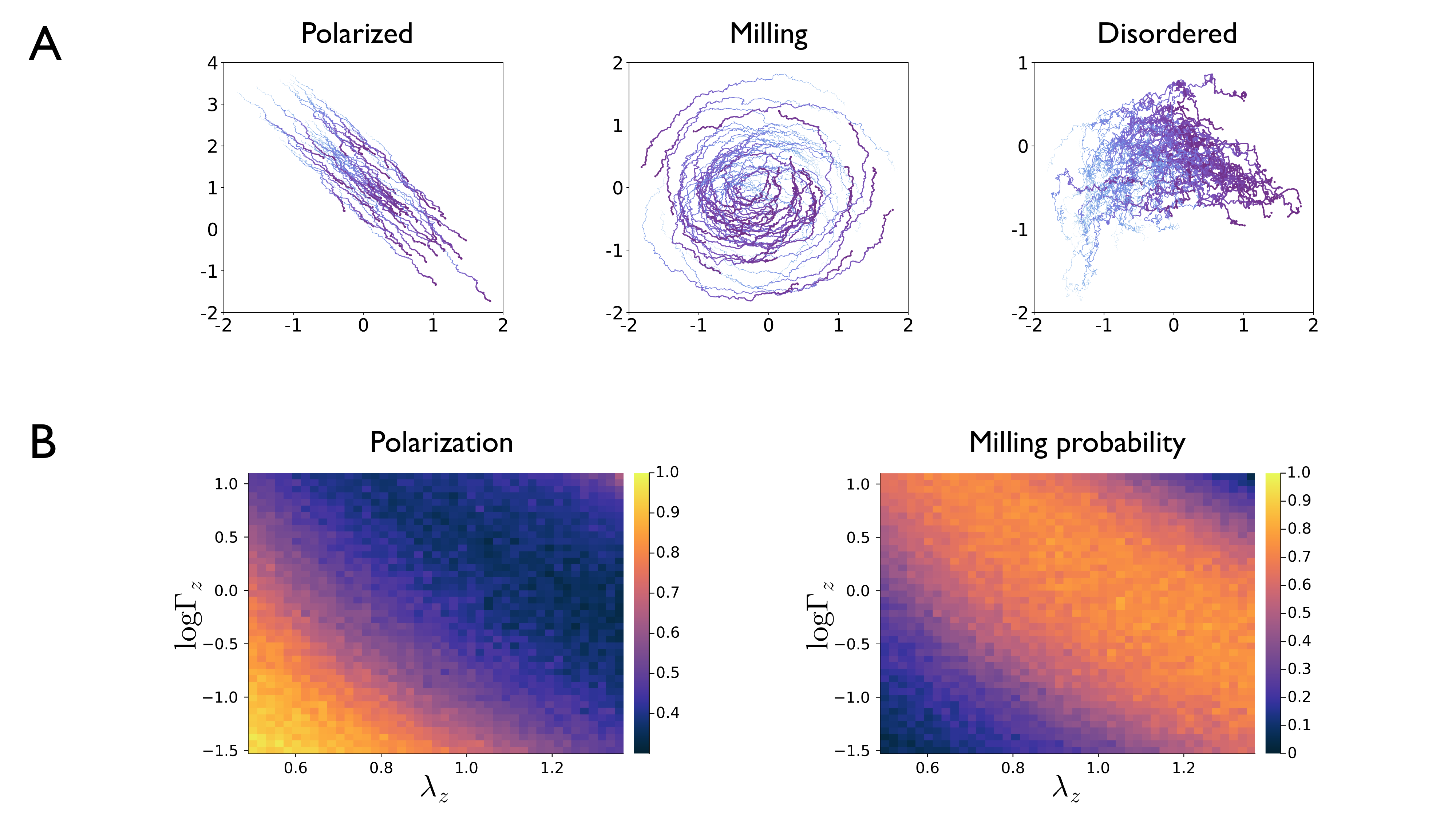}
\caption{\textbf{A}: Example snapshots of different collective states in schools of $N = 50$ active inference agents. Each line represents the trajectory of one individual, and color gradient represents time, from earliest (light blue) to latest (purple). The polarized regime in the left panel was simulated with the default parameters listed in supplementary Table E1. The milling regime (middle panel) was achieved by increasing the variance of velocity fluctuations (encoded in $\sigma^2_{z',h}$) from $0.01$ to $0.05$ (relative to the default configuration) and increasing $\lambda_z$ from $1.0$ to $1.2$. The disordered regime was achieved by increasing the sensory smoothness parameter to $2.0$ and decreasing $\eta$ from $1.0$ to $0.5$ and $\alpha$ from $0.5$ to $0.1$ (relative to the default configuration). \textbf{B}: Average polarization (left) and milling probability (right) shown as a function of the two factorized components of the sensory precision, $\Gamma_z$ (log-transformed) and $\lambda_z$. For each combination of precision parameters, we ran $500$ independent trials of `free schooling,' and then averaged the quantities of interest across trials. Each `free schooling' trial lasted $15$ seconds ($1500$ time steps with $dt = 0.01 s$); the time-averaged metrics (polarization and milling probability, respectively, were computed from the last $10$ seconds of the trial. }
\label{fig:collective_regimes_collective_properties_analysis}
\end{figure*}

\section*{Numerical results}
Given a group of active inference agents --- equipped with the generative models described in previous sections --- it is straightforward to generate trajectories of collective motion by integrating each agent's heading vector over time: $\dot{\mathbf{r}}_i = \mathbf{v}_i, i \in \{1, 2, ..., N\}$ where $N$ is the number of agents. We update all heading directions $\{ \mathbf{v}_{i} \}_{i=1}^{N}$ and beliefs $\{ \tilde{\boldsymbol{\mu}}_{i} \}_{i=1}^{N}$ in parallel via a joint gradient descent on their respective free energies:

\begin{align}
    \label{eq:coupled_gradient_descent}
    \dot{\mathbf{v}}_1 &= -\nabla_{\mathbf{v}_1} F(\tilde{\boldsymbol \mu}_1, \tilde{\mathbf{y}}_1)&  \dot{\tilde{\boldsymbol \mu}}_1 &= -\nabla_{\tilde{\boldsymbol \mu}_1} F(\tilde{\boldsymbol \mu}_1, \tilde{\mathbf{y}}_1) \notag \\
    \dot{\mathbf{v}}_2 &= -\nabla_{\mathbf{v}_2} F(\tilde{\boldsymbol \mu}_2, \tilde{\mathbf{y}}_2)&  \dot{\tilde{\boldsymbol \mu}}_2 &= -\nabla_{\tilde{\boldsymbol \mu}_2} F(\tilde{\boldsymbol \mu}_2, \tilde{\mathbf{y}}_2) \notag \\
    &\vdots& &\vdots\notag \\
    \dot{\mathbf{v}}_N &= -\nabla_{\mathbf{v}_N} F(\tilde{\boldsymbol \mu}_N, \tilde{\mathbf{y}}_N)&  \dot{\tilde{\boldsymbol \mu}}_N &= -\nabla_{\tilde{\boldsymbol \mu}_N} F(\tilde{\boldsymbol \mu}_N, \tilde{\mathbf{y}}_N)
\end{align}

For the simulation results shown here, each agent tracks the average distance $x_l$ within a total of $L = 4$ sensory sectors that each subtend $60^{\circ}$ (starting at $-120^{\circ}$ and ending at $+120^{\circ}$, relative to the focal agent's heading direction) and observe the sector-specific distances calculated using all neighbors lying within $5.0$ units of the focal agent's position. Each agent represents the vector of local distances as a generalized state with 3 orders of motion: $\tilde{\mathbf{x}} = \{\mathbf{x}, \mathbf{x}', \mathbf{x}''\}$, $\tilde{\boldsymbol{\mu}} = \{\boldsymbol{\mu}, \boldsymbol{\mu}', \boldsymbol{\mu}''\}$. Agents can observe the first and second orders of the distance $\tilde{\mathbf{y}} = \{\mathbf{y}, \mathbf{y}'\}$, i.e. the distance itself and its instantaneous rate-of-change. In the numerical results to follow, we use active inference to study the relationship between the properties of individual cognition (e.g., the parameters of agent-level generative models) and collective phenomenology.

\subsection*{Collective regimes}
Simulated groups of these distance-inferring agents display robust, cohesive collective motion (see Figure \ref{fig:collective_regimes_collective_properties_analysis}A and Supplemental Movies S1-S5). Figure \ref{fig:collective_regimes_collective_properties_analysis}A displays examples of different types of group phenomena exhibited in groups of active inference agents, whose diversity and types resemble those observed in animal groups \cite{becco2006experimental, tunstrom2013collective} and in other collective motion models \cite{vicsek1995novel, couzin2002collective, giardina2008collective}. These range from directed, coherent movement with strong inter-agent velocity correlations (`polarized motion') to group rotational patterns, like milling, which features high angular momentum around the group's center-of-mass.

\subsection*{Relating individual beliefs to collective outcomes}

In all but the most carefully constructed systems \cite{krafft2021bayesian,kaufmann2021active,heins2023spin}, the relationship between individual and collective representations is often opaque. In particular, the relationship between individual level uncertainty or `risk' and collective behavior is an open area of research. For instance, some research has indicated that increased risk-sensitivity at the level of the individual may lead to decreased risk-encoding at the collective level \cite{sosna2019individual}. Inspired by these observations, we use active inference to examine the quantitative relationship between uncertainty at the individual level and collective phenomenology.  
We begin by examining common metrics of group motion like polarization and angular momentum \cite{couzin2002collective}. In Figure \ref{fig:collective_regimes_collective_properties_analysis}B we explore how polarization and angular momentum are affected by two components of agent-level sensory uncertainty (i.e., inverse sensory precision): 1) the absolute precision that agents associate to sensory noise and 2) the autocorrelation or `smoothness' associated to that noise.

These components are encoded in each agent's observation model, which assumes generalized distance observations $\tilde{\mathbf{y}}$ are normally-distributed around the generalized state $\tilde{\mathbf{x}}$:

\begin{align}
    P(\tilde{\mathbf{y}}|\tilde{\mathbf{x}}) = \mathcal{N}(\tilde{\mathbf{y}}; \tilde{\mathbf{x}}, \tilde{\Sigma}^{\boldsymbol{z}})
\end{align}

Where we focus on the parameterization of the inverse of the covariance matrix, a.k.a., the precision matrix $\tilde{\Pi}^{\boldsymbol{z}}=\left(\tilde{\Sigma}^{\boldsymbol{z}}\right)^{-1}$. This precision matrix factorizes into two sub-matrices, one encoding the amplitude of random fluctuations $\mathbf{z}$ and one encoding their temporal smoothness,  i.e., the inverse of the covariance between different derivatives of random fluctuations (e.g., between $z$ and $z'$):

\begin{align}
    \tilde{\Pi}^{\mathbf{z}} &=  \Pi^{\mathbf{z}} \otimes \tilde{\Pi}^{z} \notag \\
    \text{where} \hspace{2.5mm}
    \Pi^{\mathbf{z}} &= \begin{bmatrix} \Gamma_{z,1} & 0  &\hdots & 0\\
    0 & \Gamma_{z,2} &  &\\
    \vdots & & \ddots \\
    0 & & & \Gamma_{z,L}
    \end{bmatrix} \\
    \tilde{\Pi}^{z} &= \begin{bmatrix} 1 & 0 \\ 0 & 2\lambda_z^2\end{bmatrix}
   \label{eq:generalized_precisions}
\end{align}

Intuitively, $\Gamma_z$ encodes the variance or amplitude that the agent associates with the noise in each of its $L$ sensory sectors $z_l$, and $\lambda_z$ encodes how ‘smooth’ the agent believes the noise is \cite{friston2007variational, parr2021computational}. A higher value of $\lambda_z$ implies that the agent believes sensory noise is more serially-correlated (e.g., random fluctuations in optical signals caused by smooth variations in refraction due to turbulence in water). Section C.1 of the \textit{SI Appendix} shows how the smoothness parameter $\lambda_z$ can be derived from a noise process with a Gaussian autocorrelation function. The consequences of this parameterization can be mapped back to the first-order prediction errors $\boldsymbol{\xi}'_{z}$ that drive action in \eqref{eq:expanded_action_rule} and \eqref{eq:multivariate_action}:

\begin{align}
    \boldsymbol{\xi}'_{z} &= \begin{bmatrix} 2\Gamma_{z,1} \lambda_z^2 (y'_{h,1} - \mu'_{h,1}) \\
    2\Gamma_{z,2} \lambda_z^2 (y'_{h,2} - \mu'_{h,2}) \\ \vdots \\ 2\Gamma_{z,L} \lambda_z^2 (y'_{h,L} - \mu'_{h,L}) \end{bmatrix}
\end{align}

Here, we have simply written the precision assigned to noise $z_{h,l}$ in a particular sensory sector as a product of the amplitude and smoothness parameters: $\pi'_{z, l} = 2\Gamma_{z,l} \lambda_z^2$.

Figure \ref{fig:collective_regimes_collective_properties_analysis}B shows how the different components (amplitude and smoothness) of the agent's beliefs about uncertainty determine group behavior, as quantified by average polarization and milling probability. Average polarization is defined here as the time average of the polarization of the group, where the polarization at a given time $p(t)$ measures the alignment of velocities of agents comprising the group \cite{couzin2002collective, buhl2006disorder}:

\begin{equation}
    \hat{p} = \frac{1}{T-t_0} \sum_{t=t_0}^{T} p(t) \hspace{10mm} 
    p(t) = \frac{1}{N} \| \sum_{i = 1}^{N} \mathbf{v}_i(t) \|
\end{equation}

Note that the time average is calculated once steady-state has been reached, where the beginning of this state is indicated by $t_0$ (for the heatmaps shown in \ref{fig:collective_regimes_collective_properties_analysis}B, we calculate these average metrics with $t_0 = 5 s$). High average polarization indicates directed, coherent group movement. The left panel of Figure \ref{fig:collective_regimes_collective_properties_analysis}B shows how $\Gamma_z$ and $\lambda_z$ contribute to the average polarization of the group. An increase in either parameter causes polarization to decrease and angular momentum to increase, reflecting the transition from directed motion to a milling regime, where the group rotates around its center of mass. We calculate the milling probability (c.f. right panel of Figure \ref{fig:collective_regimes_collective_properties_analysis}B) as the proportion of trials where the time-averaged angular momentum surpassed $0.5$. The average angular momentum can be used to quantify the degree of rotational motion, and is calculated as the time- and group-average of the individual angular momenta around the groups' center of mass $\mathbf{c}$:

\begin{equation}
    \hat{m} = \frac{1}{T-t_0} \sum_{t=t_0}^{T} m(t) \hspace{10mm} 
    m(t) = \frac{1}{N} \| \sum_{i = 1}^{N} \mathbf{r}_{ic}(t) \times \mathbf{v}_i(t) \|
\end{equation}

where $\mathbf{r}_{ic}$ is a relative position vector for agent $i$, defined as the vector pointing from the group center $\mathbf{c}$ to agent $i$'s position: $\mathbf{r}_i - \mathbf{c}$. We observed a large range of $\Gamma_z$ and $\lambda_z$ for which the milling regime (high average angular momentum) was stable (Figure 2B, right side). This stands in contrast to earlier self-propelled particular models like the original 3-zone Couzin model, where milling was only stable under a relatively limited range of parameters \cite{couzin2002collective}.

These collective changes can be understood by recalling how first-order prediction errors $\boldsymbol{\xi}'_{z}$ (and thus the velocity update) depend on $\Gamma_z$ and $\lambda_z$:

\begin{align}
    \boldsymbol{\xi}'_{z} &\propto 2\Gamma_z \lambda_z^2 \label{eq:pred_error_gammalambda_relation}
\end{align}

In practice, this means that as the group believes in more predictable (less rough) first-order sensory information $\mathbf{y}'_z$, the group as a whole is more likely to enter rotational, milling-like regimes. However, the enhancing effect of these first-order prediction errors $\boldsymbol{\xi}'_{z}$ on rotational motion is bounded; if prediction errors are over-weighted (e.g. high $\Gamma_z$ and/or $\lambda_z$), the group becomes more polarized again and likely to fragment (see \textit{SI Appendix}, Fig. E1). This fragmentation probability occurs at both low and high levels of $\Gamma_z$ and $\lambda_z$, implying that there is an optimal range of individual-level sensory precision where cohesive group behavior (whether polarized or milling) is stable. Thus, our model predicts that assuming one's sensory information is highly-precise is neither required, or in fact even desirable, for animals in order to facilitate collective motion. 

We have seen how one can use active inference to relate features of individual-level beliefs (in this case, beliefs about sensory precision) to collective patterns, focusing in the present case on common metrics for studying collective motion like polarization and the tendency to mill.

In the following sections, we move from looking at group-level patterns that occur during free movement, to studying the consequences of individual-level uncertainty for collective information-processing. We begin by investigating how collective information transfer depends on individual-level beliefs about the relative precisions associated with different types of sensory information.

\begin{figure*}
\centering
\includegraphics[width=0.8\linewidth]{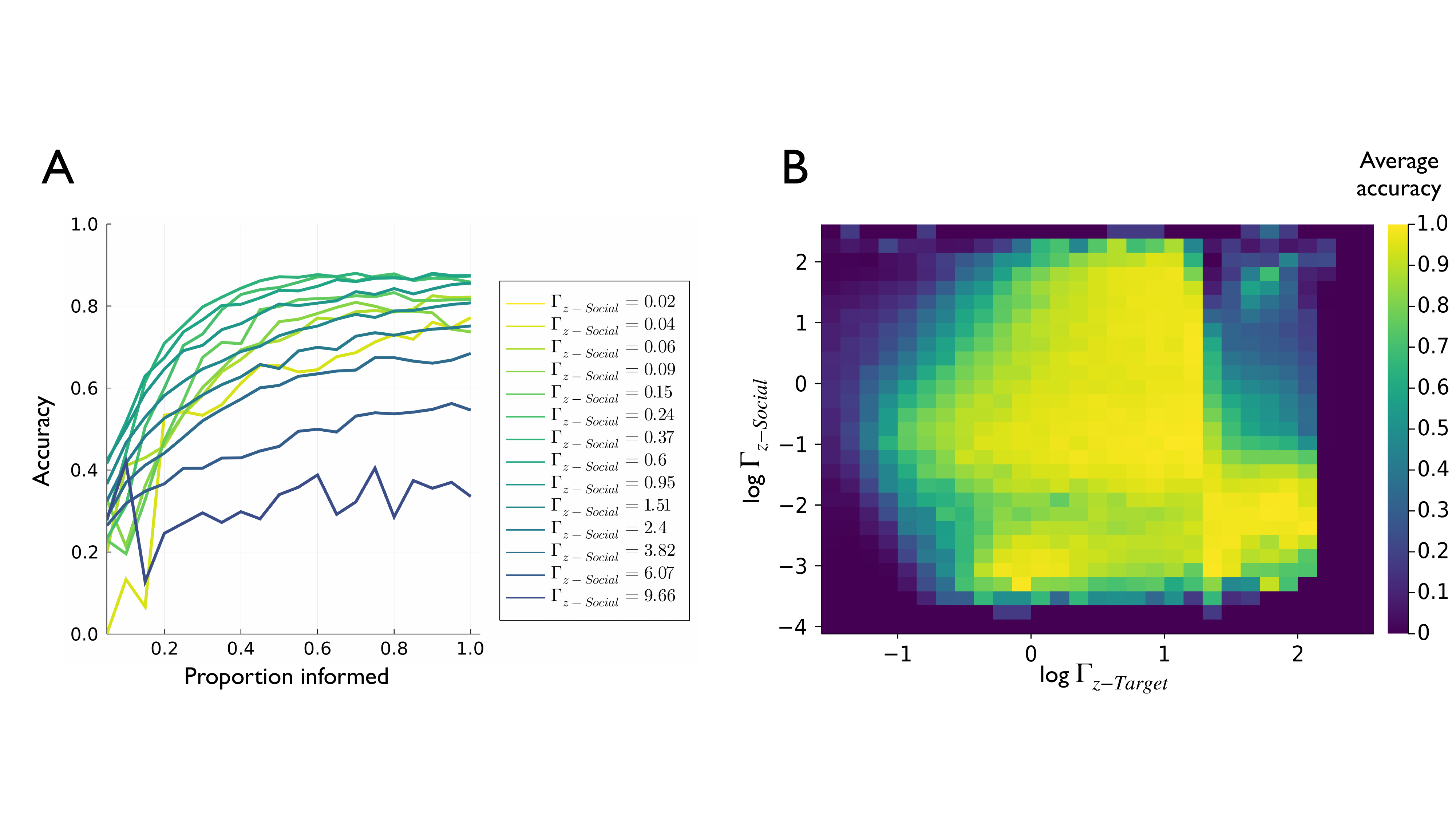}
\caption{\textbf{A}: Collective accuracy as a function of proportion informed or $p_{inf}$ for differing values of the sensory precision assigned to social observations $\Gamma_{z-\text{Social}}$. Average accuracy for each condition (combination of $p_{inf}, \Gamma_{z-\text{Social}}, \Gamma_{z-\text{Target}}$) was computed as the proportion of successful hits across $500$ trials. Here, the average accuracy is further averaged across all the values of the $\Gamma_{z-\text{Target}}$ parameter, meaning each accuracy here is computed as the average of $15000$ total trials ($500$ trials per condition $\times$ $30$ different values of $\Gamma_{z-\text{Target}}$). \textbf{B}: Collective accuracy as a function of both the social and target precisions ($\Gamma_{z-Social}, \Gamma_{z-Target}$, shown in log-scale) averaged across values of $p_{inf}$ ranging from $p_{inf} = 0.15$ to $p_{inf} = 0.40$. Each condition's accuracy was computed as the proportion of accurate decisions from $500$ trials.}
\label{fig:collective_acc_as_pinformed_gammazsoc_heatmap_accur_both_precisions}
\end{figure*}
\subsection*{Collective information transfer}

In this section, we take inspiration from the collective leadership and decision-making literature to investigate how individuals in animal groups can collectively navigate to a distant target \cite{couzin2005effective,couzin2011uninformed,strandburg2015shared, sridhar2021geometry}. This phenomenon is an example of effective leadership through collective information transfer and is remarkable for a number of reasons; one that speaks to its emergent nature, is the fact that these collective decisions are possible despite --- and indeed even because of --- the presence of uninformed individuals in the group \cite{couzin2011uninformed}. Figure \ref{fig:collective_acc_as_pinformed_gammazsoc_heatmap_accur_both_precisions}A shows that active inference agents engaged in this task reproduce a result from earlier work \cite{couzin2005effective} on the relationship between the proportion of uninformed individuals and collective accuracy. Namely, as the proportion of informed individuals increases, so does the accuracy of reaching the majority-preferred target. In the same vein as earlier sections, we also investigated the dependence of this effect, as well as the average target-reaching accuracy, on individual-level beliefs.

We operationalize the notion of an agent being `informed' (about an external target) by introducing a new latent variable to its generative model; this variable $x_{\text{target}}$ represents the distance between the informed agent's position $\mathbf{r}$ and a point-mass-like target with position vector $\mathbf{T} = [T_1, T_2]$. We thus define this new hidden state and observation as follows: $x_{\text{target}} = \| \mathbf{T} - \mathbf{r}\|$, $y_{\text{target}} = x_{\text{target}} + z_{\text{target}}$. Just like the `social' distance observations $\mathbf{y}_h$, this target distance observation $y_{\text{target}}$ represents a (potentially-noisy) observation of the true distance $x_{\text{target}}$. As before, the agent represents both the target distance $x_{\text{target}}$ and its observations $y_{\text{target}}$ using generalized coordinates of motion. Each informed agent has a dynamics model of $\tilde{x}_{\text{target}}$, whereby they assume the target-distance is driven by some drift function $f_{\text{target}}(x_{\text{target}}) = -\alpha_{t} x_{\text{target}}$ which relaxes to $0$. As with the social distances, we truncate the agent's generalized coordinates embedding of the target distance to three orders of motion and the generalized observations to two orders of motion.

Each informed agent maintains a full posterior belief $\tilde{\boldsymbol{\mu}} = (\tilde{\mu}_1, \tilde{\mu}_2, ..., \tilde{\mu}_L, \tilde{\mu}_{\text{target}})$ about the  local distances $\tilde{x}_1, \tilde{x}_2, ..., \tilde{x}_L$ as well as the target distance $\tilde{x}_{\text{target}}$.

Using identical reasoning to arrive at the action updates in \eqref{eq:expanded_action_rule} and \eqref{eq:multivariate_action}, one can augment the matrix-vector product in \eqref{eq:multivariate_action} with an extra sensorimotor contingency and prediction error that represents target-relevant information:

\begin{align}
    \frac{d\mathbf{v}}{dt} &= \tilde{\boldsymbol \xi}_{z}^{\top}\begin{bmatrix} \Delta \hat{\mathbf{R}} \\ \Delta \mathbf{T} \end{bmatrix}  \notag \\
    \Delta \mathbf{T} &=  -\nabla_{\mathbf{v}}\tilde{y}_{\text{target}} = \frac{\mathbf{T} - \mathbf{r}}{\| \mathbf{T} - \mathbf{r} \|}
    \label{eq:multivariate_action_with_target}
\end{align}

This matrix-vector product can then be seen as a weighted combination of social and target vectors, with the weights afforded to each equal to their respective precision-weighted prediction errors:

\begin{align}
    \frac{d\mathbf{v}}{dt} &= \underbrace{\xi_{\text{social}} \Delta \hat{\mathbf{R}}}_{\text{Social vector}} + \underbrace{\xi_{\text{target}} \Delta {\mathbf{T}}}_{\text{Target vector}} \label{eq:social_target_weighted_expression}
\end{align}

This expression is analogous to the velocity update in Equation (3) of Ref. \cite{couzin2005effective}, where a `preferred direction' vector is integrated into the agent's action update with some pre-determined weight. This weight is described as controlling the relative strengths of non-social vs. social information. For active inference agents, the weighting of target-relevant information emerges naturally as a precision-weighted prediction error (here represented as $\xi_{\text{target}}$), and the target-vector itself is equivalent to a sensorimotor reflex arc, that represents the agent's assumptions about how the local flow of the target distance $y'_{\text{target}}$ changes as a function of the agent's heading direction $\mathbf{v}$. An important consequence of this construction, is that, unlike in previous models where this weight is `baked-in' as a fixed parameter, the weight assigned to the target vector is dynamic, and fluctuates according to how much the agent's expectations about the target distance $\tilde{\mu}_{\text{target}}$ predict the sensed target distance $y_{\text{target}}$.

Using this new construction, we can simulate a group of active inference agents, in which some proportion $p_{inf}$ of agents represent this extra set of target-related variables as described above. To generate $\tilde{y}_{\text{target}}$ observations for these informed individuals, we placed a spatial target at a fixed distance away from the group's center-of-mass and then allowed the informed individuals to observe the generalized target distance $\tilde{y}_{\text{target}} = (y_{\text{target}}, y'_{\text{target}})$. We then integrated the collective dynamics over time and measured the accuracy with which the group was able to navigate to the target (see Materials and Methods for details). By performing hundreds of these trials for different values of $p_{inf}$, we reproduced the results of Ref. \cite{couzin2005effective} in Figure \ref{fig:collective_acc_as_pinformed_gammazsoc_heatmap_accur_both_precisions}. We see that as the number of informed individuals increases, collective accuracy increases. However, this performance gain depends on the agents` beliefs about sensory precision, which we now dissociate into two components: $\Gamma_{z\text{-Social}}$ ( the precision assigned to the social distance observations) and $\Gamma_{z\text{-Target}}$ (the precision assigned to target distance observations). By varying these two precisions independently, which respectively scale $\xi_{\text{social}}$ and $\xi_{\text{target}}$ in \eqref{eq:social_target_weighted_expression}, we can investigate the dependence of collective accuracy on the beliefs of individual agents about the uncertainty attributed to different sources of information. 

Figure \ref{fig:collective_acc_as_pinformed_gammazsoc_heatmap_accur_both_precisions}A shows the average collective accuracy as a function of $p_{inf}$, for different levels of the social distance precision $\Gamma_{z\text{Social}}$. The pattern that emerges is that the social precision, that optimizes collective decision-making, sits within a bounded range. The general effect of social precision is to essentially balance the amplification of target-relevant information throughout the school, with the need for the group to maintain cohesion. When social precision is too high, agents over-attend to social information and are not sensitive to the information provided by informed individuals; when it is too low, the group is likely to fragment and will not accurately share target-relevant information; meaning only the informed individuals will successfully reach the target. Figure \ref{fig:collective_acc_as_pinformed_gammazsoc_heatmap_accur_both_precisions}B shows that a similar optimal precision-balance exists for $\Gamma_{z\text{Target}}$. Here, we show average collective accuracy (averaged across values of $p_{inf}$ as a function of social- and target-precision. Maximizing collective accuracy appears to rely on agents balancing the sensory precision they assign to different sources of information; under the active inference model proposed here, this balancing act can be exactly formulated in terms of the variances (inverse precisions) afforded to different types of sensory cues.

\subsection*{Online plasticity through parameter learning}

The ability of groups to tune their response to changing environmental contexts, such as rapid perturbations or informational changes, is a key feature of natural collective behavior \cite{sosna2019individual,fahimipour2023wild}. However, many self-propelled particle models lack a generic way to incorporate this behavioral sensitivity \cite{couzin2005effective} and exhibit damped, `averaging'-like responses to external inputs \cite{kolpas2013spatial}. This results from classical models usually equipping individuals with fixed interaction rules and constant weights for integrating different information sources. While online weight-updating rules and evolutionary algorithms have been used to adaptively tune single-agent parameters in some cases \cite{couzin2005effective,sridhar2021geometry, bizyaeva2022nonlinear}, these approaches are often not theoretically principled and driven by specific use-cases [with notable exceptions \cite{hamann2014evolution,kaiser2022innate, gandolfi2022emergence}].

Active inference offers an account of tune-able sensitivity, using the same principle used to derive action and belief-updating in previous sections: minimizing surprise. In practice, this sensitivity emerges when we allow agents to update their generative models per se in real-time. Updating generative model parameters over time is often referred to as ``learning'' in the active inference literature \cite{friston2016active}, since it invokes the notion of updating beliefs about parameters rather than states, where parameters and states distinguish themselves by fast and slow timescales of updating, respectively. We leverage this idea to allow agents to adapt their generative models and thus adapt their behavioral rules, referring to this process as plasticity, in-line with the notion of short-term plasticity in neural circuits \cite{hennig2013theoretical}. To enable agents to update generative model parameters, we can simply augment the coupled gradient descent in \eqref{eq:coupled_gradient_descent} with an additional dynamical equation, this time by minimizing free energy with respect to model parameters, which we subsume into a set $\theta$:

\begin{align}
    \dot{\theta} &= -\nabla_{\theta} F(\tilde{\boldsymbol{\mu}}, \tilde{\boldsymbol{y}}, \theta) \label{eq:learning_rule}
\end{align}

The generative model parameters $\theta$ represent the statistical contingencies or regularities agents believes govern their sensory world; this includes the various precisions associated with sensory and process noises $\tilde{\Pi}^{z}, \tilde{\Pi}^{\omega}$ and the parameters of the dynamics and observation models, $\tilde{\mathbf{f}}, \tilde{\mathbf{g}}$. Since the free energy is a smooth function of all the generative model parameters, in theory learning can be done with respect to any parameter using procedure entailed by \eqref{eq:learning_rule}.

\begin{figure*}
\centering
\includegraphics[width=0.8\linewidth]{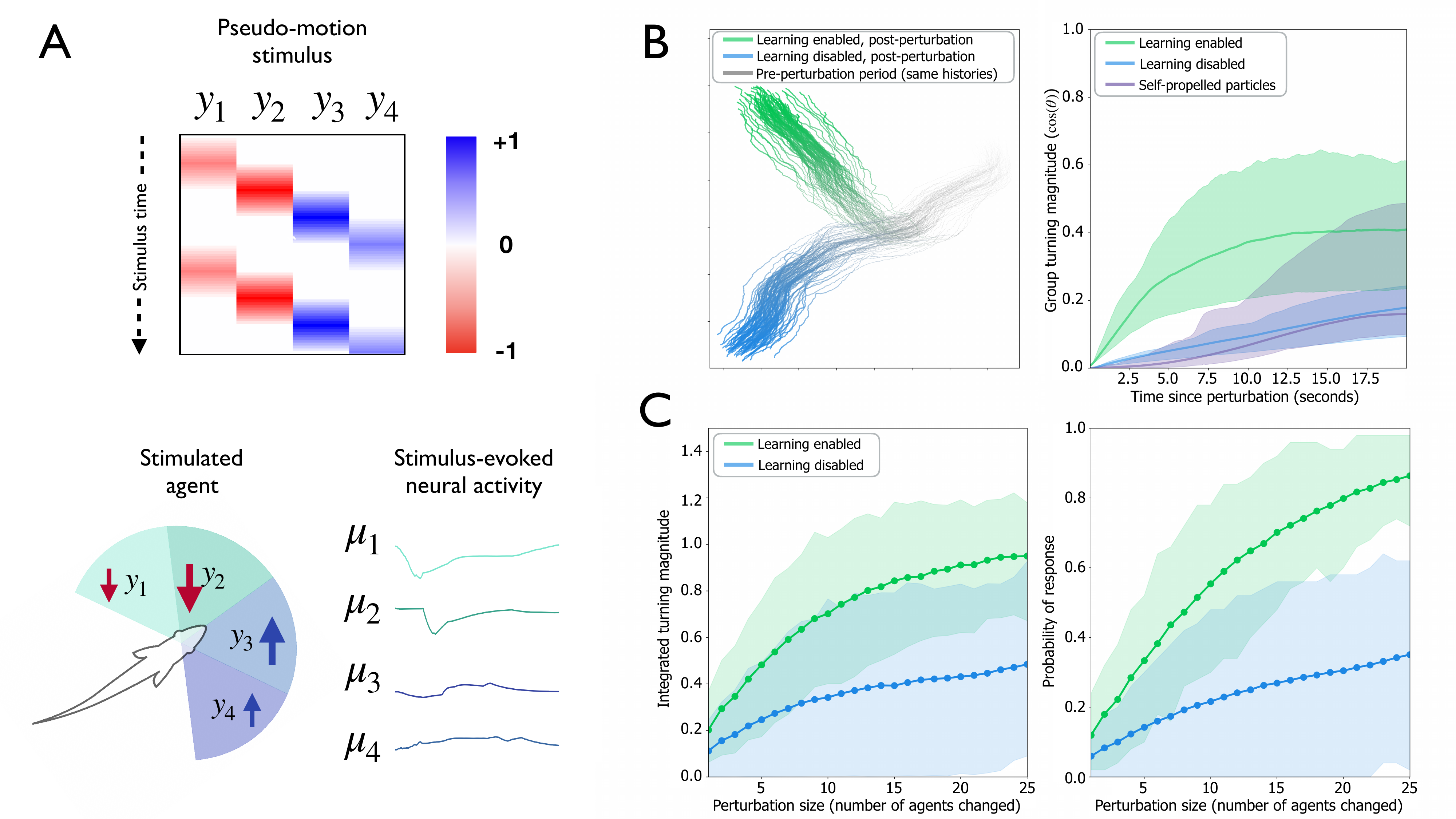}
\caption{\textbf{A}: Schematic of the sensory perturbation protocol. The `pseudo-motion' stimulus consists of repetitively perturbing the agent's sensory sectors with a moving wave of prediction errors in the agent's velocity-observation modality $\mathbf{y}'_{h}$. The top panel shows stimulus pattern as a heatmap over (amplitude over time) with two repetitions, starting from negative (red, sectors $1$ and $2$) and transitioning to positive (blue, sectors $3$ and $4$) prediction errors. The sign-switch in the stimulus (from negative to positive) mimics a moving object that first moves towards focal individual and then moves away. The temporal order of the stimulus across the sectors can be used to selectively emulate a right-moving vs. left-moving object, relative to the focal individual's heading-direction. The bottom panel shows how the stimulated agent's beliefs about the distance hidden state $\boldsymbol \mu$ changes over the course of the motion stimulus, with these beliefs being analogized to hypothetical neural activity. \textbf{B}: Response magnitude to a perturbation in presence or absence of parameter learning. Left panel: example pair of 2-D trajectories of active inference agents with matched pre-perturbation histories, in response to an individual perturbation. The ability to perform parameter-learning is left on in one stochastic realization (green) and turned off in the other (blue), following the perturbation. Right panel: initialization-averaged collective responses (group turning angle) to perturbation of active inference agents when learning is enabled or disabled. The perturbation response of a 2-zone self-propelled particle model (purple line) based on \cite{couzin2005effective} is also shown for reference. \textbf{C}: Collective response as a function of the number of perturbed individuals, comparing simulations where parameter-learning is enabled to those where it's disabled. Shown is the mean response with highest density regions (HDRs) of integrated turning magnitude within 500-1000 ms of the perturbation (left) and response probability (right) computed from $N_i = 200$ independent initializations of each condition. For each initialization, the average metric is computed across $N_r = 50$ independent realizations that were run forward from the same point in time, following a sensory prediction error perturbation (to a randomly-chosen set of perturbed agents). Response probability is computed as the proportion of independent realizations, per initialization, where the group turning rate exceeded $\pi$ radians within the first 10 seconds of the perturbation.}
\label{fig:perturbation_results}
\end{figure*}

In practice, combining parameter-learning with active inference usually implies a separation of timescales, whereby learning or plasticity occurs concurrently to state inference and action but at a slower update rate. In all the results shown here, agents update parameters an order of magnitude more slowly than they update beliefs or actions. To furnish a interpretable example of plasticity, in the simulations described here, we enabled agents to update their beliefs about the sensory smoothness parameter $\lambda_z$. We chose sensory smoothness due to its straightforward relationship to the magnitude of sensory prediction errors (c.f. the relation in \eqref{eq:pred_error_gammalambda_relation} and \textit{SI Appendix}, Section C). As agents tune $\lambda_z$ to minimize free energy, belief updating and action will at the same time become quadratically more or less responsive to sensory information.

One example of where behavioral plasticity is crucial for collective information processing is a group's ability to rapidly amplify behaviorally-relevant information, e.g., detecting the presence of a predator \cite{ward2011fast,strandburg2013visual,davidson2021collective}. To study the effect of behavioral plasticity on collective responsiveness, we perturbed single agents in groups of active inference agents while enabling or disabling online plasticity. We perturbed groups by inducing transient `phantom' prediction errors in random subsets of agents and measuring the resulting turning response of the group (see Materials and Methods for details). These prediction errors were structured (see Figure \ref{fig:perturbation_results}A) to mimic a transient visual stimulus, e.g., a loom stimulus or approaching predator \cite{harpaz2021precise}, which reliably induces a sustained turning response in the chosen individual \cite{kolpas2013spatial}. Figure \ref{fig:perturbation_results} shows the effect of enabling plasticity on the size and sensitivity of collective responses to these perturbations. Not only do plasticity-enabled groups respond more strongly to perturbations of single-agents, compared to their plasticity-disabled counterparts (\ref{fig:perturbation_results}B), but the magnitude of the collective response is also more sensitive to the size of the perturbation (\ref{fig:perturbation_results}C). As has been measured in biological collectives \cite{gomez2023fish}, the plasticity-enabled groups collectively encode the size of perturbations with higher dynamic range than plasticity-disabled controls.

The active inference framework provides a flexible and theoretically-principled approach to modeling adaptive, collective behavior with tuneable sensitivity, that eschews ad-hoc update rules or expensive evolutionary simulations. The plasticity mechanism proposed here is not limited to updating beliefs about sensory smoothness: it can be extended to update beliefs about any model parameter using the same principle. The ability to adapt generative model parameters in real-time represents a promising avenue for future research in active inference and collective behavior, and may lead to more biologically-plausible hypotheses about the mechanisms underlying adaptive responses in the natural world.

\section*{Discussion}

We have proposed active inference as a flexible, cognitively-inspired model class that can be used in the theoretical study of collective motion, as well as in empirical settings as an individual-level model of behavior. By framing behavior as the consequence of prediction-error minimization --- with respect to an individual's world model --- we offer examples of how naturalistic collective motion emerges in, where individual behavior is driven by the imperative to minimize the surprisal associated with sensory signals. Under mild distributional assumptions, this surprise is scored by an interpretable proxy; namely, prediction error. In the particular case of collective motion, a group of active inference agents equipped with a simple generative model of local social information can recover and generalize the social forces that have been the core mechanism in classical SPP models of collective motion. The active inference framework also provides a probabilistic interpretation of ad-hoc `weight' parameters that are often used in these models, in terms of the precisions that agents associate to different types of sensory information.  

We have also shown how the active inference framework can be used to characterize the relationship between generative model parameters and emergent information-processing capacities, as measured by collective information transfer and responsiveness to external perturbations. 
Active inference's generality allows us to relax the typically-static behavioral rules of SPP models, by enabling agents to flexibly tune their sensitivity to prediction errors. This is achieved via principled processes like parameter learning (i.e., `plasticity'), and can be used to model naturalistic features of collective behavior, such as the tendency to amplify salient (i.e., precise) information, that have largely evaded modelling in the SPP paradigm, except in cases where adaptation rules are explicitly introduced \cite{couzin2005effective, sridhar2021geometry}. However, when we simply allow agents to update parameters, in addition to beliefs and agents, using the principle of surprise-minimization, many hallmarks of these naturalistic behaviors can be easily obtained. 

The surprise minimization approach adopted here is both theoretically grounded in fundamental physical, cybernetic and informational principles \cite{conant1970every, friston2006free, wissner2013causal, hornischer2019structural} while also biologically-inspired, due to the scalability of the belief and action update rules, which are hypothesized to be implementable on neuronal circuits \cite{spratling2017review}. Our approach thus also harmonizes with modern `data-driven' approaches in behavioral biology, that aim to quantitatively estimate the behavioral algorithms used by different biological systems directly from experimental data \cite{calovi2014swarming, hein2020algorithmic, fahimipour2023wild}. 

By providing a flexible modeling approach that casts perception, action, and learning as manifestations of the single drive to minimize surprise, we have highlighted active inference as a toolbox for studying collective behavior in natural systems. Future work in this area could explore how the framework can be used to investigate other forms of collective behavior (not just collective motion), like multi-choice decision-making, social foraging and communication \cite{friedman2021active,albarracin2022epistemic}. The results shown in the current work serve primarily as a proof of concept: we started by writing down a specific, hypothetical active inference model of agents engaged in group movement, and then generated naturalistic behaviors by integrating the resulting equations of motion (i.e., free energy gradients) for this particular model. Taking inspiration from fields like computational psychiatry \cite{montague2012computational, smith2020recent}, we emphasize the ability to move from simple forward modelling of behavior to data-driven model inversion, whereby one hopes to infer the values of parameters that best explain empirical data (of e.g., behavioral movement data). Instead of using `force mapping' techniques to estimate social forces from behavioral measurements \cite{escobedo2020data, mudaliar2023examination}, our approach would instead frame the problem as one of computational phenotyping, where alternative generative models that a particular animal might be equipped with, could be estimated from behavioral or neural data acquired from that animal. The resulting `social forces' or interaction rules would then emerge as those behaviors that minimize surprise, relative to the generative model that best explains the animal's behavior. Both the estimation of model parameters and alternative model structures can be achieved through Bayesian model inversion and system identification methods like Bayesian model selection, averaging or reduction \cite{penny2006bayesian}. 

\section*{Materials and Methods}\label{sec:matmethods}

For all simulations we randomly initialized the positions and (unit-magnitude) velocities of $N$ particles, and integrated the equations of motion for active inference and generalized filtering using a forwards Euler-Maruyama scheme with an integration window of $\Delta t = 0.01 s$ (see \textit{SI Appendix}, Section E for details). We varied group size $N$ and the length of the simulation $T$ (in seconds) depending on the experiment. Detailed background on generalized filtering, active inference, and derivations specific to the generative model we used for collective motion can be found in the \textit{SI Appendix}, Section A. All other parameters used for simulations, unless stated otherwise, are listed in \textit{SI Appendix}, Table E.1. The code (written in JAX and Julia) used to perform simulations can be found in the following open-source repository: \href{https://github.com/conorheins/collective_motion_actinf}{https://github.com/conorheins/collective\_motion\_actinf} \cite{heins2023collective_motion}.
\subsection*{Quantifying fragmented groups}
For all experiments, we excluded trials where the group failed to maintain cohesion (or fragmented) to a sufficient degree. We deemed any given trial fragmented, when at least one individual was further than 2.0 dimensionless units away from all other individuals for at least 3 of the last 10 seconds of the trial. For the perturbation experiments, groups were excluded if this criterion was reached during the last 5 seconds of the 20 second post-perturbation period.

\subsection*{Collective information transfer experiments}
For each trial of collective target-navigation, we initialized a group of $N = 30$ agents with random positions and velocities (centered on the origin) and augmented the generative models of a fixed proportion $p_{inf}$ of the total number of agents, where $p_{inf}$ ranged from $0.05$ to $1.0$, with extra latent and observed variables representing the distance to the target with position vector $\mathbf{T}$. The distance to the target was always $10$ units from the origin. We measured collective accuracy as follows: we count a given trial as successful if the group is able to navigate to within $0.25$ units of the target without losing cohesion within $T = 15$ seconds (the length of each trial). The accuracy for a given experimental condition was then computed as the proportion of successes observed in $500$ total trials.

\subsection*{Perturbation experiments}
For the perturbation experiments, we simulated $N_i = 200$ randomly-initialized independent runs of $N=50$ agents, which we term independent initializations. We ran each initialization forward for $T=100$ seconds, a point at which metrics like average polarization, angular momentum, and median nearest-neighbor distance were highly likely to have stopped changing and fluctuate around a stationary value. Starting at $T=100$ we then split each initialization into two further sets of $N_r = 50$ parallel realizations. Each realization used a different random seed used to A) generate the action- and observation-noises; and B) select the candidate agent(s) for perturbation. Note that the splitting of seeds at $T=100$ means that each realization has an identical history up until that point. We enabled parameter learning of $\lambda_z$ in one set of realizations and we left it disabled in the other. We then perturbed random subsets of agents in both learning-enabled and -disabled realizations (2\% - 50\% of the group, i.e., 1 to 25 agents), by transiently inducing first-order prediction errors $\boldsymbol{\xi}'_z$ in the perturbed individuals. We computed the relative group turning angle after the perturbation for $20 s$ to generate the plots in Figure \ref{fig:perturbation_results}B and C.

\paragraph{Acknowledgements:} The authors would like to thank Brennan Klein, Jake Graving, Armin Bahl, Dimitrije Markovic, Thomas Parr, Pawel Romanczuk, and Manuel Baltieri for discussions during the writing of this manuscript, and Maya Polovitskaya for creating the fish schematic used in the figures.
CH and IDC acknowledge support from the Office of Naval Research Grant N0001419-1-2556, Germany's Excellence Strategy-EXC 2117-422037984 (to IDC), the Max Planck Society, the European Union's Horizon 2020 research and Innovation Programme under the Marie Skłodowska-Curie Grant agreement (to IDC; \#860949), the PathFinder European Innovation Council Work Programme (to IDC; \#101098722), and the John Templeton Foundation (to CH; \#61780).
LD is supported by the Fonds National de la Recherche, Luxembourg (Project code: 13568875) and the Engineering and Physical Sciences Research Council Centre for Doctoral Training in Mathematics of Random Systems: Analysis, Modelling and Simulation (EP/S023925/1). RPM is supported by UK Research and Innovation Future Leaders Fellowship MR/S032525/1 and the Templeton World Charity Foundation Inc. TWCF-2021-20647. KF is supported by funding for the Wellcome Centre for Human Neuroimaging (Ref: 205103/Z/16/Z), a Canada-UK Artificial Intelligence Initiative (Ref: ES/T01279X/1) and the European Union’s Horizon 2020 Framework Programme for Research and Innovation under the Specific Grant Agreement No. 945539 (Human Brain Project SGA3).

\begin{sloppypar}
\printbibliography[title={References}]
\end{sloppypar}

\clearpage
\appendix
\setcounter{figure}{0}
\setcounter{table}{0}
\setcounter{equation}{0}
\renewcommand\thefigure{\thesection.\arabic{figure}}
\renewcommand\thetable{\thesection.\arabic{table}}
\renewcommand\theequation{\thesection .\arabic{equation}}
\begin{refsection}

\section{An active inference model of collective motion}\label{sec:appendix_A}

Each agent within our model of collective motion maintains an internal model of its local environment represented by average distances to its neighbours. These distances are partitioned into $L$ sensory sectors $\mathbf{x} = {x_1, x_2, ..., x_L}$, with each agent observing noisy versions of these distances through a corresponding sensory channel $\mathbf{y} = {y_1, y_2, ..., y_L}$. Each agent estimates the hidden distance variable(s) $\mathbf{x}$ over time using its observed sensory states $\mathbf{y}$. In practice, each agent implements this through a form of variational Bayesian inference developed for continuous data-assimilation in dynamic environments called \textit{generalized filtering}, which can be seen as a variational, more flexible version of Kalman filters. This dynamic inference process entails updating posterior beliefs about $\mathbf{x}$ using a gradient descent on variational free energy. In the case of Gaussian assumptions about observation and state noise, these free energy gradients resemble a precision-weighted average of sensory and state prediction errors. This comprises the state-estimation component of active inference and is unpacked in detail in Section \ref{sec:gen_filt}. 

In addition to estimating the hidden distance variable with generalized filtering, each agent also changes its heading direction $\mathbf{v}$ in order to minimize the same variational free energy functional. When the agent's model of the distance dynamics is strongly `biased' by a prior belief that the steady-state value of the distance variable(s) $\tilde{\mathbf{x}}$ hovers around a particular value $\boldsymbol{\eta}$, then agents will change their heading in a way that appears like they `want' to maintain this target distance between them and their neighbours. Concretely, this means they move closer to neighbors when the sensed distance $\mathbf{y}$ is larger than expected, and move away from neighbors when $\mathbf{y}$ is smaller than expected. 

This symmetry between belief updating and action, as both following the gradients of the same loss function, is what theoretically distinguishes active inference from other continuous control schemes, which often use different objectives for estimation and control. In the following sections we detail the processes of state-estimation and action under active inference.

\subsection{Generalized filtering overview}
\label{sec:gen_filt}

Agents estimate hidden states $x$ as the variational solution to a Bayesian inference problem; they achieve this in practice using an online-filtering algorithm known as generalized filtering \cite{friston2010generalised,friston2008hierarchical}. Generalized filtering is a generic Bayesian filtering scheme for non-linear state-space models formulated in generalized coordinates of motion \cite{balaji2011bayesian}. It subsumes, as special cases, variational filtering \cite{friston2008variational}, dynamic expectation maximization \cite{friston2008DEMvariational} and generalized predictive coding \cite{buckley2017free}. This inversion scheme relies on a simple dynamical generative specification of hidden states $x$ and how they relate to observations $y$. The generative model starts by postulating that the time evolution of a variable $x$ is given by a stochastic differential equation with the following form:

\begin{align}
    \frac{dx_t}{dt} &= f(x_t) + \omega_t
\label{eq:f_zeroth_order_equation}
\end{align}

where $f$ is some deterministic flow function (i.e., a vector field) that depends on the current state $x_t$, and $\omega_t$ is a (smooth) additive Gaussian noise process. Under generalized filtering, we successively differentiate \eqref{eq:f_zeroth_order_equation}, to finesse the difficult computation of the \textit{paths} or trajectories of $x_t$ locally in time, by instead focusing on the much easier problem of computing the serial derivatives of $x_t$. This allows one to express a local trajectory of $\vec{x} = \{x_t, x_{t+1}, \hdots, x_{t+T}\}$ in terms of the derivatives of $x_t$, i.e., $\tilde{x}_t=(x_t', x_t'', x_t''', \hdots,x^{[n]}_t,\hdots )$, where $x^{[n]}_t := \frac{d^n}{dt^n} x_t$. We used the notation $\tilde{x}_t$ to denote a vector of these higher orders of motion at time $t$, a representation known as \textit{generalized coordinates}. The equivalence between generalized coordinates and paths locally in time follows from Taylor's theorem, where the path of $x$ around some time $t$ can be expressed as a combination of its higher order derivatives:

\begin{align}
    x_{t+h} &= x_t^{[0]} + \sum_{n = 1}^{\infty}\frac{x^{[n]}}{n!}h^{n} \label{eq:taylors_theorem}
\end{align}

Note that the (local in time) equality between a path $\vec{x}$ and its Taylor series only holds when the sample paths of $x_t$ are analytic functions, which itself requires $f$ to be analytic and the noise process $\omega_t$ to be analytic (in particular non-white noise fluctuations) \cite{friston2022free}. Successively differentiating the base equation in \eqref{eq:f_zeroth_order_equation} (and ignoring contributions of the flow of order higher than one) yields a series of stochastic differential equations that describe the evolution of each order of motion $x_t^{[n]}$ as depending on its own state and the $n$\textsuperscript{th} derivative of the noise \cite{balaji2011bayesian}:

\begin{align}
        \dot{x} &= f(x) + \omega \notag \\
        \dot{x}' &= f_{x} x' + \omega '\notag \\
        \dot{x}'' &= f_{x} x'' + \omega '' \notag \\
        &\vdots \notag \\
       \Rightarrow D \tilde{x} 
       &= \tilde{f} + \tilde{\omega} \notag
\end{align}

where, following the notation used in \cite{friston2008hierarchical, friston2010generalised, balaji2011bayesian}, we use the notation $f_x$ for the Jacobian (i.e., matrix of first order partial derivatives) of the flow function $f$ evaluated at $x$, i.e., $Jf(x)$, 
and omit the time variable from our notation for conciseness.
Note that the above construction assumes a local linearization of $f$ around $x$, in the sense that it ignores the contribution of higher order derivatives of the flow \cite{balaji2011bayesian}. The $D$ is the time derivative operator in generalised coordinates, with identity matrices along the first leading (block) diagonal and $\tilde{f}, \tilde{\omega}$ are the generalized flow function and generalized noises, respectively:

\begin{equation}
    D = \begin{bmatrix} 0 & I & \\
        & \ddots & \ddots & \\
        & & \ddots & I \\
        & & & 0
        \end{bmatrix} \hspace{10mm} \tilde{f} = \begin{bmatrix} f( x^{[0]}) \\
        f_x x^{[1]} \\
        \vdots \\
        f_x x^{[n]}
        \end{bmatrix}  \hspace{10mm}  \tilde{\omega} = \begin{bmatrix}  \omega^{[0]} \\
        \omega^{[1]} \\
        \vdots \\
        \omega^{[n]}
        \end{bmatrix} \notag \\
\end{equation}

Here, $n$ is some chosen order at which to truncate the derivatives. This truncation means that the Taylor expansion of a path $\vec{x}$ in \eqref{eq:taylors_theorem} is rendered an approximation -- valid locally in time. Having specified a dynamics over $x$ (and its reformulation in generalized coordinates), we are in a position to specify the \textit{observation model}. In generalized filtering, the generative model of state dynamics is supplemented with an observation model that maps hidden states $x$ to their sensory consequences $y$ via some (differentiable) sensory map $g(x)$ and additive Gaussian smooth fluctuations $z$:

\begin{align}
    y_t &= g(x_t) + z_t
\label{eq:g_zeroth_order_equation}
\end{align}

Like the states, we can similarly express observations in generalized coordinates by successively differentiating \eqref{eq:g_zeroth_order_equation} to obtain a similar single expression for the generalized observation equation:

\begin{align}
        y &= g(x) + z \notag \\
        y' &= g_{x} x' + z' \notag \\
        y'' &= g_{x} x'' + z'' \notag \\
        &\vdots \notag \\
        \Rightarrow\tilde{y} &= \tilde{g} + \tilde{z} \notag
\end{align}

where here the $i$\textsuperscript{th} motion of observations $y^{[i]}$ is not a function of itself but rather that of the motion of the (generalized) hidden states $x^{[i]}$ and fluctuations $z^{[i]}$. In other words, the motion of observations tracks the simultaneous motion of the states, subject to any nonlinearities in the sensory map $g$ and the motion of the noise $z$. Given Gaussian assumptions on the generalised noises $\tilde{\omega}$ and $\tilde{z}$, we can then write down the full hidden state and observation model $p(\tilde{y}, \tilde{x})$ in terms of Gaussian densities:

\begin{align}
   D\tilde{x}&= \tilde{f} + \tilde{\omega} & \tilde{\omega} &\sim \mathcal N(\tilde{\omega}; \mathbf{0}, \tilde{\Sigma}^{\omega})\notag \\ \tilde{y} &= \tilde{g} + \tilde{z} & \tilde{z} &\sim \mathcal N(\tilde{z}; \mathbf{0}, \tilde{\Sigma}^{z}) \notag \\
    \implies p(\tilde{y}, \tilde{x}) &= p(\tilde{y}|\tilde{x})p(D\tilde{x}|\tilde{x}) \notag \\ &= \mathcal N(\tilde{y}; \tilde{g}, \tilde{\Sigma}^{z})\mathcal N(D\tilde{x}; \tilde{f}, \tilde{\Sigma}^{\omega}) \label{eq:gen_filt_GM}
\end{align}

This Gaussian specification of the generative model licenses efficient, online update rules for the sufficient statistics of approximate posterior beliefs that track the expected value of the generalised hidden state $\tilde{x}$. This relies on a simple expression for the variational free energy of this state-space model; as we will see in the following sections, this not only enables efficient state estimation (a.k.a, updating beliefs about hidden states $\tilde{x}$), but also algorithms for inferring generative model parameters.

\subsection{State estimation under generalized filtering}
Generalized filtering relies on optimizing posterior beliefs in order to minimize  \textit{variational free energy} $F$, an upper bound on the \textit{surprise} associated with observations $y$ under some generative model $m$: 

\begin{equation}
    F \geq \underbrace{-\ln p(y;m)}_{\textrm{surprise}} 
\end{equation}

where the model $m$ defines a joint distribution over observations and latent variables $p(y, \vartheta)$. The latent variables themselves $\vartheta$ are often split into hidden states $x$ and parameters $\theta$. Exact Bayesian inference entails obtaining the posterior distribution over latent variables $p(\vartheta | y)$, which can be expressed using Bayes rule:

\begin{align}
    p(\vartheta | y) &= \frac{p(y, \vartheta)}{p(y)} \label{eq:bayes_rule} \\\
    p(y) &\triangleq \int p(y, \vartheta) d \vartheta \label{eq:marginal_likelihood}
\end{align}

where hereafter we leave out the dependence on the model $m$.

In order to compute the posterior exactly, one has to compute the marginal probability of observations $p(y)$, also known as the marginal likelihood or model evidence. Computing the marginal likelihood is often intractable or difficult in practice, motivating the introduction of the variational bound, the free energy $F$, also known as the (negative) evidence lower-bound or ELBO. This can be shown by writing $F$ as the Kullback-Leibler divergence between some "variational" distribution $q(\vartheta; \nu)$ over latent variables with parameters $\nu$ and the true posterior $p(\vartheta|y)$:

\begin{align}
    F &= \mathbb{E}_{q}\left[ \ln q(\vartheta) - \ln p(y, \vartheta)\right] \notag\\\ 
    &= D_{KL}\left(q(\vartheta; \nu) || p(\vartheta|y)\right)\underbrace{-\ln p(y)}_{\text{surprise}} \label{KL_plus_surprise} \\\
    \implies F &\geq -\ln p(y) \label{eq:F_upper_bound_surprise}
\end{align}

The upper bound holds because the Kullback-Leibler divergence is always non-negative $ D_{KL}(p||q) \geq 0$. Intuitively, as the variational distribution $q(\vartheta; \nu)$ better approximates the true posterior distribution $p(\vartheta|y)$, where the (in)accuracy of the approximation is measured by the KL divergence, then the tighter the free energy bounds the surprise. This decomposition also makes clear why minimizing $F$ with respect to variational parameters $\nu$ is a way to update the variational distribution $q$ to approximate the true posterior $p(\vartheta|y)$. The variational distribution is thus often referred to as an approximate posterior, where the exact posterior obtained by applying Bayesian rule as in Equation \eqref{eq:bayes_rule} corresponds to the variational posterior that minimises $F$.

Now we turn to deriving the Laplace-approximation to the variational free energy (VFE) for the Gaussian state-space models used in generalised filtering. The Laplace approximation is an analytically tractable way to approximate the true posterior with a Gaussian distribution, which simplifies inference to an online filtering algorithm that corresponds to minimizing a sum of squared prediction errors.

Recall that our goal is to perform inference on the latent variables $\vartheta$ by optimizing an approximate posterior distribution $q(\vartheta; \nu)$. In our case, we let $\vartheta = \{x, \theta\}$ where $x$ are hidden states and $\theta$ encompass other generative model parameters (e.g., hyperparameters of the generative model like $\tilde{f}, \tilde{g}, \tilde{\Sigma}^{z}, \tilde{\Sigma}^{\omega}$). For now we focus on inference over hidden states $x$ and treat parameter inference later. Under the Laplace approximation we use a Gaussian distribution for the approximate posterior distribution $q(x; \nu)$:

\begin{align}
    q(x; \nu) &= \mathcal{N}(x; \underbrace{\mu, \Sigma^{\nu}}_{\nu}) \label{eq:variational_gaussian}
\end{align}

where the variational parameters $\nu$ are comprised of the sufficient statistics of a Gaussian distribution: the mean $\mu$ and covariance $\Sigma^{\nu}$. We add the subscript $\nu$ to the variational variance to distinguish it from generative model covariances, e.g. $\tilde{\Sigma}^{z}, \tilde{\Sigma}^{\omega}$.

We can now arrive at a more specific expression for the variational free energy using the Gaussian form of the variational distribution. We start by decomposing the free energy into the sum of an expected energy term and a (negative) entropy,  where the energy is defined as the negative log joint density over states and observations: $-\ln p(x, y)$ and the negative entropy is that of the variational posterior i.e., $\mathbb{E}_q[\ln q(x; \nu)]$: 

\begin{align}
    F = \mathbb{E}_{q}\left[-\ln p(x, y)\right] - \frac{1}{2}\left[\ln \left|\Sigma\right|+d \ln 2 \pi e\right]  \label{eq:LAP_step2}
\end{align}

where $d$ is the dimensionality of $x$ and the full term on the right follows from the entropy of a multivariate Gaussian: $\operatorname{H}[\mathcal N(x ; \mu, \Sigma)] = \frac{1}{2}\left[\ln \left|\Sigma\right|+d \ln 2 \pi e\right]$.

Additional assumptions allow one to further simplify the expected energy term $\mathbb{E}_{q}\left[-\ln p(x, y)\right]$; namely, if we assume that the posterior is tightly peaked around the mean $\mu$ and that $p(x, y)$ is twice-differentiable in $x$, we can motivate a 2\textsuperscript{nd}-order Taylor expansion of the expected energy term around its mode, i.e. when $x = \mu$: 
\begin{align}
    \mathbb{E}_{q}\left[-\ln p(x, y)\right] &\approx \mathbb{E}_{q}\left[ -\ln p(\mu, y) - \nabla_x \ln p(x, y)\biggr\rvert_{x=\mu}(x - \mu) -  \frac{1}{2}(x - \mu)^\top\nabla^2_x \ln p(x, y)\biggr\rvert_{x=\mu}(x - \mu) \right] \notag \\\
    &=  -\ln p(\mu, y) - \frac{1}{2}\operatorname{tr}\left(\Sigma\nabla^2_x \ln p(x, y)\biggr\rvert_{x=\mu} \right)\label{eq:LAP_step3}
\end{align}


Combining this approximation of the expected energy with the remaining terms in the variational free energy, we can now write the full expression of the Laplace-approximated free energy $F_{L}$:

\begin{align}
    F_{L} = -\ln p(\mu, y) - \frac{1}{2}\operatorname{tr}\left(\Sigma \nabla^2_x \ln p(x, y)\biggr\rvert_{x=\mu}\right) - \frac{1}{2} \left(\ln \left|\Sigma\right|+d \ln 2 \pi e\right) \label{eq:LAP_step4}
\end{align}

A useful feature of this expression is that the optimal variational covariance $\Sigma^{\nu}$ can obtained by setting the derivative of $F_{L}$ with respect to the covariance $\Sigma$ equal to 0 and solving for $\Sigma$, i.e. finding the values of the covariance that minimize the $F_{L}$:

\begin{align}
    \frac{\partial F_{L}}{\partial \Sigma} = 0 &\iff \Sigma^{\nu} = -\left(\nabla^2_x \ln p(x, y)\biggr\rvert_{x=\mu}\right)^{-1}
    \label{eq:optimal_variance}
\end{align}

i.e., the optimal variance of the variational distribution is the curvature of the Laplace energy around its mode. Substituting this expression back into the full free energy, we can then write an expression that only depends on the mean vector $\mu$ of the variational density, since the variatonal variance $\Sigma^{\nu}$ is now expressed as a function of the mean:

\begin{align}
    F_{L} &= -\ln p(\mu, y) + \frac{1}{2}\underbrace{\operatorname{tr}\left(\Sigma^{\nu} (\Sigma^{\nu})^{-1} \right)}_{=d} - \frac{1}{2} \left(\ln \left|\Sigma^{\nu}\right|+d \ln 2 \pi e \right) \notag \\
    &=  -\ln p(\mu, y) - \frac{1}{2} \left(\ln \left|\Sigma^{\nu}\right|+d \ln 2 \pi \right)\label{eq:LAP_final}
\end{align}

This means that the Laplace approximation to the variational free energy is a function of only the variational mean $\mu$ and sensory observations $y$, because the variational variance $\Sigma^{\nu}$ is itself a function of $\mu$. Belief updating then consists in minimizing the Laplace-approximated free energy $F_{L}$ with respect to $\mu$:

\begin{align}
    \dot{\mu} \propto -\nabla_{\mu} F_{L}(\mu, y) \label{eq:grad_descent_VFE}
\end{align}

Which consists of descending the gradient of the energy and log determinant terms with respect to $\mu$. When the generative model $p(x, y)$ is Gaussian, the energy term is quadratic in $\mu$ and $y$. This means that its gradient can be written in terms of precision-weighted \textit{prediction errors}, which score the difference between the expected observations (given the current value of $\mu$) and the actual observations $y$. This notion of using prediction errors to estimate hidden quantities is also known as predictive coding \cite{friston2005theory, friston2009predictive, huang2011predictive}. The log determinant term --- in all of our cases of interest --- turns out to have a vanishing gradient with respect to $\mu$. In summary, the free energy gradient is a sum of precision-weighted \textit{prediction errors}, and $\mu$ evolves to minimize those prediction errors.

To illustrate this, we take the simplest example --- that of a linear, static, joint Gaussian generative model, where the prior over hidden states $p(x)$ is a Gaussian density with mean $\eta$ and covariance $\Sigma^{\omega}$, and the observation model $p(x|y)$ is a Gaussian density with mean $g(x)$, which is some linear function of the hidden state:

\begin{equation}
\label{eq: Gaussian gen model}
\begin{split}
y \sim \mathcal N(g(x), \Sigma^{z}), \quad x \sim \mathcal N(\eta, \Sigma^{\omega}).
\end{split}
\end{equation}

For this linear Gaussian generative model, the variational mean $\mu$ only influences the expected energy term of $F_{L}$, because the optimal covariance $\Sigma^{\nu}$ is independent of $\mu$. Thus we ignore the constant entropy term and write out the energy as a sum of precision-weighted prediction errors:

\begin{align}
    -\ln  p(\mu, y) &= -\ln p(y|\mu) - \ln p(\mu) \notag \\
    &= \frac{1}{2} \left[\varepsilon_{z}^{T}\Pi^{z}\varepsilon_{z} + \varepsilon_{\omega}^{T}\Pi^{\omega}\varepsilon_{\omega}
    \right]  \notag \\
    \text{where } &\Pi^{z} = (\Sigma^{z})^{-1}, \hspace{2mm} \Pi^{\omega} = (\Sigma^{\omega})^{-1} \notag \\
    \text{and } &\varepsilon_{z} = y - g(\mu), \hspace{2mm} \varepsilon_{\omega} = \mu - \eta
\end{align}

We can write out gradients of this quadratic energy function to yield the update equation for the means $\mu$ as in \eqref{eq:grad_descent_VFE}, and see that $\mu$ changes as a precision-weighted sum of `sensory` and `model' prediction errors (up to additive constants):

\begin{align}
    \dot{\mu} &= -\nabla_{\mu} F_{L}(\mu, y) \notag \\
    &= -\nabla_{\mu} \left[\frac{1}{2} \left(\varepsilon_{z}^{T}\Pi^{z}\varepsilon_{z} + \varepsilon_{\omega}^{T}\Pi^{\omega}\varepsilon_{\omega}\right)\right] \notag \\
    &= -\left[
    (g_\mu)^T\Pi^{z}\varepsilon_{z} + \Pi^{\omega}\varepsilon_{\omega}\right] \label{eq:grad_quadratic_VFE}
\end{align}

Note that the variational means only depend on the terms of $F_{L}$ containing $\varepsilon_z$ and $\varepsilon_\omega$, so that the update reduces to a gradient descent on a sum of squared prediction errors. This belief update scheme illustrates the key principles of predictive coding under the Laplace approximation: conditional means, denoted as $\mu$, change as a function of precision-weighted prediction errors. The concept of precision-weighting in belief updating is intuitive: if the generative model attributes higher variance to sensory fluctuations as compared to state variance (i.e., $\Pi^{z} < \Pi^{\omega}$), then sensory data is relatively unreliable and consequently makes a smaller impact on posterior beliefs. Therefore, the adjustment to the posterior mean $\mu$ in \eqref{eq:grad_quadratic_VFE} is primarily influenced by the state prediction error term $\Pi^{\omega}\varepsilon_{\omega}$ or the prior. Conversely, when sensory information is allocated higher precision (lower variance) relative to prior beliefs (i.e., $\Pi^{z} > \Pi^{\omega}$), belief updates will strongly rely on sensory data.

We apply the above steps to derive the Laplace-approximated free energy with a Gaussian posterior $q(x; \nu)$ to the dynamical generative model in \eqref{eq:gen_filt_GM},which is constructed from Gaussian densities. Note that we use the tilde notation to now indicate that all variables are vectors of generalised coordinates, e.g., $\tilde{y}, \tilde{x}$, etc. Proceeding exactly as above, the Laplace free energy is a sum of the energy, a log determinant term, and a constant term, i.e. \eqref{eq:LAP_final}. Unlike in the linear Gaussian case, the potential non-linearity in the flows make that the log determinant term varies with respect to $\mu$. However, as it turns out its gradient is approximately zero under the local linear approximation. Therefore, the only term that matters in the free energy as its gradient does not vanish is the energy term. In summary, we write:

\begin{align}
    F_L &\propto \tilde{\varepsilon}_{z}^{T}\tilde{\Pi}^{z}\tilde{\varepsilon}_{z} + \tilde{\varepsilon}_{\omega}^{T}\tilde{\Pi}^{\omega}\tilde{\varepsilon}_{\omega} \label{eq:VFE_just_PEs_dyn}\\
    \tilde{\varepsilon}_{z} &\triangleq \tilde{y} - \tilde{g} \notag \\ \tilde{\varepsilon}_{\omega} &\triangleq D\tilde{\mu} - \tilde{f}
    \notag
\end{align}

Here, the so-called `generalised errors' $\tilde{\varepsilon}_{z}$ and $\tilde{\varepsilon}_{\omega}$ encapsulate sensory and state prediction errors across orders of motion. Belief updating is again performed using a gradient descent on free energy, but the dynamic nature of inference necessitates an additional 'motion' term:

\begin{align}
\frac{d \tilde{\mu}}{d t} &= D\tilde{\mu} - \nabla_{\tilde{\mu}}F_L \notag \\
&= D\tilde{\mu} +{ g}_{ \tilde{\mu}}^\top \tilde{\xi}_{z} + { f}_{ \tilde{\mu}}^\top \tilde{\xi}_{\omega} - D^\top\tilde{\xi}_{\omega} \notag \\
\textrm{where} \hspace{2mm} \tilde{\xi}_{z} &= \tilde{\Pi}^{z}\tilde{\varepsilon}_{z} \notag \\
\tilde{\xi}_{\omega} &= \tilde{\Pi}^{\omega}\tilde{\varepsilon}_{\omega}
\label{eq:PC_update_dynamic_gm}
\end{align}

The additional term $D\tilde{\mu}$ places the gradient descent within the context of the expected movement of the conditional means $\tilde{\mu}$, and hence of the free energy minimum. This concept has been referred to as 'gradient descent in a moving frame of reference' \cite{friston2010generalised}. This implies that free energy minimization does not occur when the beliefs cease moving, but rather when the belief update rate $\frac{d \tilde{\mu}}{dt}$ is identical to the beliefs about the motion itself $D\tilde{\mu}$, in other words when $\frac{\partial F}{\partial \tilde{\mu}} = 0 \iff D\tilde{\mu} = \frac{d \tilde{\mu}}{d t}$. This additional temporal correction proves beneficial in a dynamic data assimilation regime, where incoming observations are integrated online with beliefs that are evolving according to their own prior dynamics \cite{friston2010generalised}.

\subsection{Active inference for continuous control}
\label{sec:act_inf}

Active inference casts action or control as issuing from the same process of free energy minimization as used for state estimation; the only difference is that we now have an additional set of variables, actions $a$, that can be changed to minimize free energy as well. The update equation for actions $a$ closely resembles that used to update the variational mean $\mu$, i.e., a gradient descent on the (Laplace-encoded) variational free energy:

\begin{align}
    \frac{da}{dt} &= -\frac{\partial F_{L}(\mu, y(a))}{\partial a} \notag \\
    &= -\frac{\partial F_L}{\partial y(a)}\frac{\partial y(a)}{\partial a} 
    \label{eq:action_update_suppl} 
\end{align}

where we have now introduced a dependence between of observations $y$ on actions $a$. This allows us to express the free energy gradient with respect to action as the product of the derivative of the free energy with respect to observations $\nabla_y F_L(\mu, y(a))$ and the derivative of the function mapping from actions to observations $\frac{\partial y(a)}{\partial a}$. The free energy gradient with respect to observations is exactly the sensory prediction error $\nabla_y F_L(\mu, y(a)) = \xi_z = \Pi (y - g(x))$. This assumed dependence of observations on actions underwrites the notion that active inference agents cannot directly measure how their actions affect hidden states, but may only do so via their sensory consequences. This has been speculated to explain the architecture of descending motor pathways in corticospinal systems, where motor commands are `unpacked' into proprioceptive predictions at the level of spinal circuits and other lower motor nuclei. Action is thus realized by minimizing proprioceptive prediction errors via classical reflex arcs \cite{adams2013predictions}. The reflex arc term $\frac{\partial y(a)}{\partial a}$ of \eqref{eq:action_update_suppl} is analogous to a forward model in motor control \cite{friston2011optimal}, because it reflects the agent's implicit assumptions about how the agent's own actions lead to their (anticipated) sensory consequences. This sort of update rule leads active inference agents to minimize sensory prediction errors via these `baked-in' sensorimotor contingencies. In this way active inference has been referred to as `action by self-fulfilling prophecy' \cite{buckley2017free}. In other words, the agent generates top-down expectations of `preferred' sensory inputs, which then generates prediction errors which can then be suppressed through low-level motoric reflexes \cite{adams2013predictions}.

\subsection{Filtering and control for a self-propelled particle}
\label{sec:act_inf}

Having derived a routine for state estimation and action through a generalized gradient flow on the Laplace-approximated variational free energy $F_{L}$, we can now apply this to the simulation of collective motion. In what follows, we write down a sufficient generative model for a single self-propelled agent and unpack the corresponding free energy gradients (\eqref{eq:VFE_just_PEs_dyn} and \eqref{eq:action_update_suppl}) using the structure and parameters of the chosen generative model. In this section we unpack the per-agent generative model of local distances described in the main text and demonstrate how a more parametric, unconstrained version of social forces are reproduced by minimizing free energy with respect to the distance-tracking generative model.

\subsubsection{A generalised filter for local distances and their time evolution}

As described in the main text, each agent represents a  an $L$-dimensional vector $\mathbf{x}$ where $\mathbf{x} = (x_1, x_2, ..., x_L)$.\footnote{We use the bold notation $\mathbf{x}$ to represent a vector-valued variable} The agent not only represents the instantaneous value (or `position') of $\mathbf{x}$ but also its generalized motion, which we truncate at 3\textsuperscript{rd} order:

\begin{align}
        \dot{\mathbf{x}} &= f(\mathbf{x}) + \boldsymbol{\omega} \notag \\
        \dot{\mathbf{x}}' &= f_{\mathbf{x}} \mathbf{x}' + \boldsymbol{\omega} '\notag \\
        \dot{\mathbf{x}}'' &= f_{\mathbf{x}} \mathbf{x}'' + \boldsymbol{\omega}'' \notag \\
       \Rightarrow D \tilde{\mathbf{x}} &= \tilde{\mathbf{f}} + \tilde{\boldsymbol{\omega}} \notag
\end{align}

The flow at the first order $f$ is a linear dynamical system with drift matrix $A$ and fixed point with value $\boldsymbol{\eta}$:

\begin{align}
    f(\mathbf{x}) &= -\mathbf{A}(\mathbf{x} - \boldsymbol{\eta})
\end{align}

The eigenvalues of the $L 
\times L$  matrix $\mathbf{A}$ determine the rate at which the hidden states $\mathbf{x}$ are assumed to relax to their expected value of $\boldsymbol{\eta}$. In general, this matrix can be parameterized arbitrarily to encode different kinds of linear couplings among the different hidden states $x_1, x_2, ..., x_L$. In the present work we parameterize $\mathbf{A}$ simply as a diagonal matrix with a single diagonal value $\alpha>0$, which can also be expressed as an $\alpha$-scaled version of the identity matrix $L \times L$ identity matrix $I_{L}$:

\begin{align}
    \mathbf{A} &=  -\alpha I_L
\end{align}

In combination with the amplitude of random fluctuations $\Sigma^{\omega}$, $\alpha$ determines how quickly the hidden states relax to their mean value of $\boldsymbol{\eta}$.\footnote{Heuristically, it is an exponential decay rate.} The generalised flow function $\tilde{\mathbf{f}}$ can thus be written as a linear function of the generalised state $\tilde{\mathbf{x}}$: 

\begin{align}
    \tilde{\mathbf{f}} = \begin{bmatrix} f(\mathbf{x}) \\
    f_\mathbf{x} \mathbf{x}' \\
    f_\mathbf{x} \mathbf{x}''
    \end{bmatrix} &= - \begin{bmatrix} \mathbf{A} & \mathbf{0} & \mathbf{0} \\
        \mathbf{0} & \mathbf{A} & \mathbf{0}\\
        \mathbf{0} & \mathbf{0} & \mathbf{A}
        \end{bmatrix} \begin{bmatrix} \mathbf{x} - \boldsymbol{\eta} \\
    \mathbf{x}'\\
    \mathbf{x}''
    \end{bmatrix} \notag \\
    &=  \begin{bmatrix} -\alpha I_L & \mathbf{0} & \mathbf{0} \\
        \mathbf{0} & -\alpha I_L & \mathbf{0}\\
        \mathbf{0} & \mathbf{0} & -\alpha I_L
        \end{bmatrix} \begin{bmatrix} \mathbf{x} - \boldsymbol{\eta} \\
    \mathbf{x}'\\
    \mathbf{x}''
    \end{bmatrix}  = -\alpha \begin{bmatrix} \mathbf{x} - \boldsymbol{\eta} \\
    \mathbf{x}'\\
    \mathbf{x}''
    \end{bmatrix} \label{eq:flow_function_linear}
\end{align}

where $\mathbf{0}$ are $L \times L$ matrices of zeros. We assume a multivariate Gaussian form for the generalized noises $\tilde{\boldsymbol{\omega}}$, meaning the density over the generalized motion $D\tilde{\mathbf{x}}$ is a  Gaussian density, which we hereafter refer to as the `dynamics model' or `dynamical prior':

\begin{align}
    P(D\tilde{\mathbf{x}}|\tilde{\mathbf{x}}) &= \mathcal{N}(D\tilde{\mathbf{x}}; \tilde{\mathbf{f}}, \tilde{\Sigma}^{\boldsymbol{\omega}}) 
\end{align}

Consistent with the block diagonal form of the generalised flow function $\tilde{\mathbf{f}}$, we also assume the covariance of the generalized noises $\tilde{\Sigma}^{\boldsymbol{\omega}}$ factorizes into a Kronecker product of `spatial' and `temporal' covariance matrices, i.e.,

\begin{align}
    \tilde{\Sigma}^{\boldsymbol{\omega}} &=  \Sigma^{\boldsymbol{\omega}} \otimes \tilde{\Sigma}^{\omega}
   \label{eq:generalized_precisions}
\end{align}

where the spatial covariance $\Sigma^{\boldsymbol{\omega}}$ (note the bold superscript $\boldsymbol{\omega}$) represents covariance between $L$ noise processes at the zero-th order $\boldsymbol{\omega}^{[0]}$, i.e., $\Sigma^{\boldsymbol{\omega}}= \mathbb{E}[\boldsymbol{\omega}^{[0]}\otimes \boldsymbol{\omega}^{[0]}]$, and $\tilde{\Sigma}^{\omega}$ encodes covariance between different derivatives of the first order noise, i.e., $\forall {m, n}: \left(\tilde{\Sigma}^{\omega}\right)_{nm}=\mathbb{E}[\omega^{[n]}\cdot \omega^{[m]}]$. The entries of this covariance matrix can be written in terms of the derivatives of the autocorrelation function of the random fluctuations evaluated at lag  $0$, $\rho(0)$:

\begin{align}
    \rho(h) &\triangleq (\Sigma^{\boldsymbol{\omega}})^{-1} \mathbb{E}[\omega^{[0]}(\tau)\cdot \omega^{[0]}(\tau + h)]
    \notag \\
    \Rightarrow\tilde{\Sigma}^{\omega} &= \begin{bmatrix}
    1 & 0 & \ddot{\rho}(0) & \\
    0 & -\ddot{\rho}(0) & 0 & \\
    \ddot{\rho}(0) & 0 & \ddot{\ddot{\rho}}(0) & \\
    & & & \ddots
    \end{bmatrix}
\end{align}

The checkerboard structure in the matrix reflects the fact that fluctuations at the first order are orthogonal to their motion (first derivative), but anti-correlated with their 2\textsuperscript{nd}, 4\textsuperscript{th}, ..., etc. derivatives. A derivation of the temporal covariance matrix from the autocorrelation function of the first-order fluctuations can be found in Appendix A.5.3 of \cite{parr2022active}. 
In the generative models of our agents, we assume a Gaussian autocorrelation function with "smoothness" parameter $\lambda_{\omega}$, which yields a simple parameterization of $\tilde{\Sigma}^{\omega}$:

\begin{align}
\rho(h)&=e^{-\frac{h}{2\lambda_{\omega}}^2}\\
    \Rightarrow\tilde{\Sigma}^{\omega} &= \begin{bmatrix} 1 & 0 & -\frac{1}{2\lambda^2_{\omega}}  &\hdots \\
    0 & \frac{1}{2\lambda^2_{\omega}} &0 & \\
    -\frac{1}{2\lambda^2_{\omega}} & 0 & \frac{3}{4\lambda^4_{\omega}} & \\
    \vdots & & & \ddots
    \end{bmatrix}
\end{align}

A higher value of $\lambda_{\omega}$ dampens the variance of the generalised fluctuations at higher orders of differentiation. The correspondence of increasing $\lambda_{\omega}$ to an increasingly-autocorrelated process at the first order becomes intuitive once we consider the case of standard white noise, i.e., the derivative of the Wiener process, whose higher orders of motion have infinite variance (the state of the process at a given time changes infinitely quickly). 
This ability to handle differentiable noise goes beyond the usual Markovian assumptions made in standard state space models (e.g., Kalman-Bucy filters), which assume that the driving noise is white.

We parameterize the $L \times L$ spatial covariance $\Sigma^{\boldsymbol{\omega}}$ through its precision matrix $\Pi^{\boldsymbol{\omega}}$, as a diagonal matrix whose entries are given by a single precision (inverse variance) $\Gamma_{\omega}$:

\begin{align}
\Sigma^{\boldsymbol{\omega}} &= (\Pi^{\boldsymbol{\omega}})^{-1} = \begin{bmatrix} \Gamma_{\omega} & 0 & 0  &\hdots \\
    0 & \Gamma_{\omega} &0 & \\
    0 & 0 & \Gamma_{\omega} & \\
    \vdots & & & \ddots
    \end{bmatrix}^{-1} \label{eq:spatial_covariance}
\end{align}

The observation likelihood describes sensory observations $\mathbf{y} = \{y_1, y_2, ..., y_L\}$ as noise-perturbed copies of the hidden states $\mathbf{x}$. We truncate generalized observations at second order, i.e., agents can sense the first order hidden state $\mathbf{x}$ and its motion $\mathbf{x}'$:

\begin{align}
    \mathbf{y} &= \mathbf{x} + \mathbf{z} \notag \\
    \mathbf{y}' &= \mathbf{x}' + \mathbf{z}' \label{eq:observation_model_real}
\end{align}

This can be equivalently expressed as a linear function $\tilde{\mathbf{g}}$ of the full generalised state $\tilde{\mathbf{x}} = \{\mathbf{x}, \mathbf{x}', \mathbf{x}''\}$, where $\tilde{\mathbf{g}}$ represents multiplication with a non-invertible matrix that discards acceleration information $\mathbf{x}''$:

\begin{align}
    \tilde{\mathbf{y}} &= \tilde{\mathbf{g}} + \tilde{\mathbf{z}} \notag \\ \begin{bmatrix} \mathbf{y} \\ \mathbf{y}' \end{bmatrix} &= \begin{bmatrix} I_L & \mathbf{0} & \mathbf{0} \\ \mathbf{0} & I_L & \mathbf{0}  \end{bmatrix} \begin{bmatrix} \mathbf{x} \\ \mathbf{x}' \\ \mathbf{x}'' \end{bmatrix} + \begin{bmatrix} \mathbf{z} \\ \mathbf{z}'\end{bmatrix}
\end{align}
 
We leverage the same assumptions about the sensory noises $\tilde{\mathbf{z}}$ as we did for the state noises $\tilde{\mathbf{\omega}}$ to end up with the following multivariate Gaussian form for the observation model:

\begin{align}
    p(\tilde{\mathbf{y}}|\tilde{\mathbf{x}}) &= \mathcal N(\tilde{\mathbf{y}}; \tilde{\mathbf{g}}, \tilde{\Sigma}^{\mathbf{z}})
\end{align}

We parameterize the likelihood model's sensory noises $\tilde{\mathbf{z}}$ identically to the state noises $\tilde{\boldsymbol{\omega}}$, namely using a spatial precision parameter $\Gamma_z$ and temporal smoothness parameter $\lambda_z$.

Having specified the dynamics and observation models in terms of Gaussian distributions, we can write out the full generative model as a joint Gaussian density over (generalized) hidden states and observations. We can furthermore define an approximate posterior over the hidden states $\tilde{\mathbf{x}}$ that has a multivariate Gaussian form $Q(\tilde{\mathbf{x}}) =\mathcal N(\tilde{\mathbf{x}}; \tilde{\boldsymbol{\mu}}; \Sigma^{\nu})$, which can be summarized entirely in terms of its posterior mean vector $\tilde{\boldsymbol{\mu}}$, due to the fact that under the Laplace approximation the variational covariance depends directly on the mean. From here, we can define the Laplace-approximated variational free energy for this generative model as proportional to a sum of squared prediction errors:

\begin{align}
    \label{eq:laplace_vfe_specific}
    p(\tilde{\mathbf{y}}, \tilde{\mathbf{x}}) &= p(\tilde{\mathbf{y}}|\tilde{\mathbf{x}})p(D\tilde{\mathbf{x}}|\tilde{\mathbf{x}}) \notag \\
    &= \mathcal N(\tilde{\mathbf{y}}; \tilde{\mathbf{g}}, \tilde{\Sigma}^{\mathbf{z}})\mathcal N(D\tilde{\mathbf{x}}; \tilde{\mathbf\mathbf{f}}, \tilde{\Sigma}^{\boldsymbol{\omega}}) \\
    F_{L} &= \frac{1}{2}\left[\tilde{\boldsymbol{\varepsilon}}_z^{\top}\tilde{\Pi}^{\boldsymbol{z}}\tilde{\boldsymbol{\varepsilon}}_z + \tilde{\boldsymbol{\varepsilon}}_{\omega}^{\top}\tilde{\Pi}^{\boldsymbol{\omega}}\tilde{\boldsymbol{\varepsilon}}_{\omega}- \ln\left( |\tilde{\Pi}^{\mathbf{z}}| |\tilde{\Pi}^{\boldsymbol{\omega}}| |\Pi^{\nu}|\right) + 3L \ln 2\pi\right] \notag \\
    \textrm{where  } \Pi^{\nu} &\triangleq (\Sigma^{\nu})^{-1} \notag \\
    \tilde{\boldsymbol{\varepsilon}}_{z} &=  \tilde{\mathbf{y}} - \tilde{\mathbf{g}}({\tilde{\boldsymbol{\mu}}}) = \begin{bmatrix} \mathbf{y} - \boldsymbol{\mu} \\ \mathbf{y}' - \boldsymbol{\mu}'\end{bmatrix}, \hspace{2mm} \tilde{\boldsymbol{\varepsilon}}_{\omega} = D\tilde{\boldsymbol{\mu}} - \tilde{\mathbf{f}}(\tilde{\boldsymbol{\mu}}) = \begin{bmatrix} \boldsymbol{\mu}' + \alpha(\boldsymbol{\mu} - \boldsymbol{\eta}) \\ \boldsymbol{\mu}'' + \alpha\boldsymbol{\mu}' \\ \alpha \boldsymbol{\mu}''\end{bmatrix}
\end{align}

where the sensory prediction errors $\tilde{\boldsymbol{\varepsilon}}_z$ score the difference between the generalized observations $\mathbf{y}, \mathbf{y}'$ and their expected values $\boldsymbol{\mu}, \boldsymbol{\mu}'$, and the model or process prediction errors $\tilde{\boldsymbol{\varepsilon}}_{\omega}$ score the difference between the motion of the generalized means $D\tilde{\boldsymbol{\mu}}$ and their expected motion $\tilde{\mathbf{f}}(\tilde{\boldsymbol{\mu}})$, which has been expanded above using the linear form of the flow function detailed in \eqref{eq:flow_function_linear}. Note that here, due to the Laplace approximation, the generative model's expectation functions $\tilde{\mathbf{g}}, \tilde{\mathbf{f}}$ are evaluated at the variational mean $\tilde{\boldsymbol{\mu}}$, rendering the variational beliefs a moving point-estimate of the hidden states $\tilde{\mathbf{x}}$.

Filtering consists of updating $\tilde{\boldsymbol{\mu}}$ as a generalized gradient flow on this energy functional $F_L$ as in \eqref{eq:PC_update_dynamic_gm_suppl}. To be explicit, below we expand these free energy gradients using the particular forms of $\tilde{\mathbf{g}}, \tilde{\mathbf{f}}$ used by our self-propelled particle agent:

\begin{align}
\frac{d \tilde{\boldsymbol{\mu}}}{d t} &= D\tilde{\boldsymbol{\mu}} - \nabla_{\tilde{\boldsymbol{\mu}}}F_L \notag \\
&= D\tilde{\boldsymbol{\mu}} + \nabla_ {\tilde{\boldsymbol{\mu}}}\tilde{\mathbf{g}}^\top\tilde{\boldsymbol{\xi}}_{z} + \nabla_{\tilde{\boldsymbol{\mu}}} \tilde{\mathbf{f}}^\top \tilde{\boldsymbol{\xi}}_{\omega} - D^\top\tilde{\boldsymbol{\xi}}_{\omega} \notag \\
\textrm{where} \hspace{2mm} \tilde{\boldsymbol{\xi}}_{z} &= \tilde{\Pi}^{\mathbf{z}}\begin{bmatrix} \mathbf{y} - \boldsymbol{\mu} \\ \mathbf{y}' - \boldsymbol{\mu}'\end{bmatrix} \notag \\
\tilde{\boldsymbol{\xi}}_{\omega} &= \tilde{\Pi}^{\boldsymbol{\omega}}\begin{bmatrix} \boldsymbol{\mu}' + \alpha(\boldsymbol{\mu} - \boldsymbol{\eta}) \\ \boldsymbol{\mu}'' + \alpha\boldsymbol{\mu}' \\ \alpha \boldsymbol{\mu}''\end{bmatrix} \notag \\
\nabla_ {\tilde{\boldsymbol{\mu}}}\tilde{\mathbf{g}} &= \begin{bmatrix} I_L & \mathbf{0} & \mathbf{0} \\ \mathbf{0} & I_L & \mathbf{0}  \end{bmatrix}, \hspace{2mm} \nabla_{\tilde{\boldsymbol{\mu}}} \tilde{\mathbf{f}} =  \begin{bmatrix} -\alpha I_L & \mathbf{0} & \mathbf{0} \\
        \mathbf{0} & -\alpha I_L & \mathbf{0}\\
        \mathbf{0} & \mathbf{0} & -\alpha I_L
        \end{bmatrix}
\label{eq:filtering_equations_specific}
\end{align}

This sort of filtering scheme means that the agent's beliefs $\tilde{\boldsymbol{\mu}}$ will evolve as a moving average of incoming sensory data $\tilde{\mathbf{y}}$ subject to a dynamical bias or "drag", which is a consequence of the latent belief that hidden states $\mathbf{x}$ continuously relax towards a fixed point at $\boldsymbol{\eta}$. Specifically, the beliefs are constantly pulled closer to the data in order to minimize sensory prediction errors $\tilde{\boldsymbol{\xi}}_z$; however, this process itself incurs state prediction errors $\tilde{\boldsymbol{\xi}}_{\omega}$ that will pull the beliefs back towards the fixed point. This constant tug of war between sensory and process prediction errors can be shifted disproportionately in one direction by adjusting the relative precisions of the likelihood vs. dynamical models, respectively. If the process precision $\tilde{\Pi}^{\boldsymbol{\omega}}$ is high relative to the observation precision $\tilde{\Pi}^{\mathbf{z}}$, then the beliefs will tend to their expected fixed point of $\boldsymbol{\eta}$. A similar enhancement of prior bias can be achieved by increasing the drift rate $\alpha$ of the dynamics model, which increases the force driving $\boldsymbol{\mu}$ towards $\boldsymbol{\eta}$  --- this was the approach taken in \cite{baltieri2019pid}, for example.

Note that when numerically integrating the differential equation in \eqref{eq:filtering_equations_specific} with a forwards Euler scheme, one uses a finite number of iterations to update the variational means $\tilde{\boldsymbol{\mu}}$, which we term $n_{\text{InferIter}}$, and a step-size $\kappa_{\mu}$ which scales the size of the increment to $\tilde{\boldsymbol{\mu}}$ \cite{buckley2017free}. In all simulations shown here, we set $n_{\text{InferIter}} = 1, \kappa_{\mu} = 0.1$ (see Table \ref{table:tableD1} for details).

\subsubsection{Closing the loop with observations and action}

In order to interpret the random variables of the generative model as representing behaviorally-relevant features of an agent's world, we now turn to specfiying the \textit{generative process}, i.e., the actual physics of the world that our self-propelled particle agents will inhabit. In this section we detail how the observations $\tilde{\mathbf{y}}$ for a single agent are generated from the positions and velocities of other active inference agents, and how actions can be generated through \textit{active inference}, which in this contexts means changing continuous control variables using a gradient descent on the same free energy used to derive the belief update equations of the previous section.

We now shift our perspective to that of a single agent, hereafter referred to as the \textit{focal individual} or \textit{focal agent}, and specify how its sensory data $\tilde{\mathbf{y}}$ are generated. We start by describing univariate hidden states and corresponding observations, where the true hidden variable is an average nearest-neighbor distance $x_{h}$. We add the $h$ subscript to distinguish these `real' variables (hidden states, observations, noise terms) from their representations in the generative model (e.g., $\tilde{x}$, $\tilde{y}$). 

We indicate the focal individual with index $i$; so the agent $i$-relative hidden state $x_{h, i}$ denotes the average nearest-neighbor distance from the perspective of agent $i$. This average distance $x_{h, i}$ is calculated from the $K$ neighbors that form the interaction set $N_{in}$ of the $i$\textsuperscript{th} focal individual. How to define the interaction set $N_{in}$ is a choice to make in each simulation, but for the case of recapitulating classical, distance-dependent social forces models, we define $N_{in}$ as those neighbors that are within a fixed distance $R_0$ of the focal individual's position:

\begin{align}
    x_{h,i} &\triangleq \frac{1}{K}\sum_{j\in N_{in}} \| \Delta\mathbf{r}_{ij} \| \notag \\\
    \textrm{where} 
    \hspace{2mm} N_{in}&\triangleq \left\{j \neq i \, :\,\| \Delta \mathbf{r}_{ij} \| \leq R_0 \right\} \label{h_definition}\\\
    K &\triangleq \vert N_{in}\vert \notag \\\
     \Delta\mathbf{r}_{ij} &\triangleq  \mathbf{r}_j - \mathbf{r}_i \label{eq:definition_of_x_h}
\end{align}

An additional filter on $N_{in}$ that is common to self-propelled particle models, is to only include neighbors that subtend some angular extent (also known as a `vision cone' or `visual field') relative to the focal agent's velocity vector $\mathbf{v}_i$. This is the approach taken in \cite{couzin2002collective}, for instance, and in the simulations examined in the main text we do the same.

The vector $\mathbf{r}_{i}$ denotes the 2-D coordinate of the focal agent, and $\mathbf{r}_{j}$ is that of neighbor $j$. $\mathbf{r}_{ij}$ thus represents the relative displacement vector of neighbour $j$, from the perspective of the focal agent $i$.

We also define the first temporal derivative of the local average distance $x'_{h,i}$:

\begin{align}
    \tilde{x}_{h, i} &\triangleq ( x_{h, i}, x'_{h,i} ) \notag \\
    x'_{h,i} &\triangleq \frac{dx_{h, i}}{dt} = \nabla_{\mathbf{r}_i}x_{h,i}\cdot \mathbf{v}_i + \sum_{j\in N_{in}}\left(\nabla_{\mathbf{r}_j}x_{h,i} \cdot \mathbf{v}_j\right)\label{eq:generalised_hidden_states_GP}
\end{align}

where $\mathbf{v}_j$ is the velocity or heading vector of neighbour $j$. The expression in \eqref{eq:generalised_hidden_states_GP} means that we can compute the first derivative or velocity of the distance $x'_{h,i}$ as a function of the positions and velocities of all agents, as opposed to some discrete-time approximation, e.g., $x'_{h,i} \approx \frac{x_{h,i}(t + \Delta t) - x_{h,i}(t)}{\Delta t}$ for some small $\Delta t$. Note that this expression for $x'_{h,i}$ assumes a local linearization of $x_{h,i}$ at the radius defined by $R_0$, i.e., this linearization will be a poor predictor of the actual change in the state $x_{h,i}(t + \Delta t) - x_{h,i}(t)$ when neighbors are instantaneously leaving or entering the interaction set $N_{in}$. Observations $\tilde{y}_{h,i}$ are perturbed versions of the hidden states with additive generalised fluctuations $\tilde{z}_{h,i}$:

\begin{align}
    y_{h,i} &= x_{h,i} + z_{h,i} \notag \\
    y'_{h,i} &= x'_{h,i} + z'_{h,i} \notag \\
    \textrm{where} \hspace{5mm} p(\tilde{z}_{h,i}) &= N(\tilde{z}_{h,i}; \mathbf{0}, \tilde{\Sigma}_{z,h})
\end{align}

In all simulations we parameterize the $\tilde{z}_{h,i}$ as independent Gaussian variables, i.e.,

\begin{align}
\tilde{\Sigma}_{z,h} &= \begin{bmatrix} \sigma^2_{z,h} & 0 \\ 0 & \sigma^2_{z',h}\end{bmatrix}
\end{align}

where the two variances $\sigma^2_{z,h}$ and $\sigma^2_{z',h}$ can be set independently. The `perception' step of our active inference process proceeds by providing these observations to the filtering equations in \eqref{eq:filtering_equations_specific}. The result is that posterior means $\tilde{\mu}$ appears to track $\tilde{x}_{h,i}$ over time, while additionally estimating its higher-order motion (acceleration) via $\mu'''$. 

Finally, we now furnish a scheme for updating actions by mapping the control variables $a$ and sensorimotor contingency terms of  \eqref{eq:action_update_suppl} to the case of our distance-tracking self-propelled agent.

We let actions be identifiable with the heading vector $\mathbf{v}_i$ of the focal individual, i.e., $a = \mathbf{v}_i$. For the simulations presented in the current paper, we always asserted that this heading have unit magnitude, but in general this constraint is not necessary.

Given this definition of actions, we can unpack the sensorimotor contingency term $\frac{\partial y(a)}{\partial a}$ that appeared in the active inference control equation of \eqref{eq:action_update_suppl}, now letting $a = \mathbf{v}$ and turning partial derivatives into Jacobians to account for vectorial nature of actions (being a velocity in 2-D) and observations (being comprised of two generalized coordinates):

\begin{align}
    \frac{d\mathbf{v}_i}{dt}
    &= -\nabla_{\mathbf{v}_i}\tilde{y}_{h,i}(\mathbf{v}_i) ^{\top} \nabla_{\tilde{y}_{h,i}(\mathbf{v}_i)}F_L 
\label{eq:velocity_update_active_inference} 
\end{align}

Note here that observations $\tilde{y}_{h,i}$ are a function of actions; this is because observations are a linear function of hidden states, which themselves are linear in the velocity vector of the focal individual $\mathbf{v}_i$ via the relation in \eqref{eq:generalised_hidden_states_GP}. Importantly, however, the distance observation $y_{h,i}$ does not directly depend on the $\mathbf{v}_i$ --- only the distance velocity $y'_{h,i}$ does. This means the sensorimotor contingency in \eqref{eq:velocity_update_active_inference} is comprised of non-zero partial derivatives only for $y'_{h,i}$:

\begin{align}
    \nabla_{\mathbf{v}_i}\tilde{y}_{h,i}(\mathbf{v}_i) &= \begin{bmatrix} \nabla_{\mathbf{v}_i}y_{h,i}(\mathbf{v}_i)\\\nabla_{\mathbf{v}_i}y'_{h,i}(\mathbf{v}_i)\end{bmatrix} = \begin{bmatrix} \mathbf{0} \\ \nabla_{\mathbf{r}_i}x_{h,i}\end{bmatrix} \label{eq:sensorimotor_contingency_velocity}
\end{align}

This has an important consequence for action, when we consider the form of the second part of the action update in \eqref{eq:velocity_update_active_inference}, the free energy gradient term $\nabla_{\tilde{y}_{h,i}}F_L$:

\begin{align}
    \nabla_{\tilde{y}_{h,i}}F_L &= \tilde{\xi}_z = \tilde{\Pi}_{z} \tilde{\varepsilon}_z = \begin{bmatrix}\Gamma_z(y_{h,i} - \mu) \\ 2\Gamma_z\lambda_z^2(y'_{h,i} - \mu')\end{bmatrix}
\end{align}

The free energy gradient with respect to observations is simply the generalized (precision-weighted) sensory error $\tilde{\xi}_z$, which we have written in terms of the observations $\tilde{y}_{h,i}$, posterior beliefs $\tilde{\mu}$ and precision parameters $\Gamma_z, \lambda_z$. The sparse form of the sensorimotor contingency in \eqref{eq:sensorimotor_contingency_velocity} means that the 0\textsuperscript{th}-order prediction error $\xi_z$ will have no effect on behavior and only the velocity prediction errors $\xi'_z$ will be relevant for the update to $\mathbf{v}_i$, i.e.,

\begin{align}
    \frac{d\mathbf{v}_i}{dt} &=  -\left(\xi_z \underbrace{\nabla_{\mathbf{v}_i}y_{h,i}(\mathbf{v}_i)}_{=\mathbf{0}} + \xi'_z \nabla_{\mathbf{v}_i}y'_{h,i}(\mathbf{v}_i)\right)    \notag \\
    &= -\xi'_z \nabla_{\mathbf{r}_i}x_{h,i} \notag \\
    &=  2\Gamma_z\lambda_z^2(y'_{h,i} - \mu') \Delta\hat{\mathbf{r}} \notag \\
    \textrm{where} \hspace{2mm} \Delta \hat{\mathbf{r}} &= \frac{1}{K}\sum_{j\in N_{in}}\frac{\Delta \mathbf{r}_{ij}}{\| \Delta \mathbf{r}_{ij} \|}
\end{align}

Note that, as for the inference update in \eqref{eq:filtering_equations_specific}, we update $\mathbf{v}_i$ using a fixed number of action iterations $n_{\text{ActionIter}}$ and step-size $\kappa_a$, where here we set $n_{\text{ActionIter}} = 1, \kappa_a = 0.1$. This action update equation has a few key implications for the behavior of active inference agents equipped with this type of generative model, and its relationship to `classical' self-propelled particle models like the Couzin-Aoki model and the Reynolds or BOIDS model \cite{Aoki1982, couzin2002collective, reynolds1987flocks}. The first is the fact that the sensorimotor contingency is identical to the `social force' vector used to drive interactions in self-propelled particle models $\Delta\hat{\mathbf{r}}$; this the average of the vectors pointing from each neighbor in the interacting set to the focal agent's position $\mathbf{r}_i$.  The sign of the precision-weighted prediction error $\epsilon'_z$ determines whether the social force is attractive (pointing towards other agents) or repulsive (pointing away from other agents). Secondly, the fact that actions only depend on velocity observations, rather than state observations, means that agents will adjust their heading according to how the (sensed) distance is instantaneously changing (its velocity), rather than its value. This lends action a predictive, anticipatory power and accounts for why we observe robust polarized motion in the absence of an explicit alignment term like in classic self-propelled particle models \cite{vicsek1995novel, couzin2002collective}. The alignment-like forces emerges from the fact that the velocity vectors of other agents $v_{j}, j \in N_{in}$ are integrated into the computation of $y'_{h,i}$ via the relation in the second line of \eqref{eq:generalised_hidden_states_GP}. 

One of the defining features of other self-propelled particle models like the Couzin-Aoki model \cite{Aoki1982, couzin2002collective} is the presence and priorization of interaction zones. The two main zones used in these models, and which on their own are sufficient for group cohesion, are a narrow repulsion zone defined by some radius $r_{r}$ and a wider attraction zone with radius $r_{a}$, where $r_{a} > r_{r}$. Neighboring agents within the repulsive radius exert repulsive forces on the focal agent, while those beyond the repulsion radius but within the attraction zone exert attractive forces, where the difference between attraction and repulsion is given by the sign of the force vector $\Delta \hat{\mathbf{r}}$. The active inference model leads to an effective notion of zones, but rather than being explicitly encoded, these zones emerge through the fixed-point attractor $\eta$ parameterizing the generative model's dynamics model $\mathbf{f}$. This is made clear when we examine the precision-weighted prediction error $\xi'_z$, which itself is a function of velocity observations $y'_{h,i}$ and velocity beliefs $\mu'$. Consider the limiting case of when inference is strongly biased by the dynamics model $\mathbf{f}$ (i.e., in the case that $\Gamma_\omega > \Gamma_z$ or large $\alpha$); the generalised beliefs $\tilde{\mu}$ will be strongly drawn to the setpoint $\eta$ of the dynamics prior, i.e.,

\begin{align}
    \tilde{\mu} = \begin{bmatrix} \mu \\ \mu' \\ \mu''\end{bmatrix} &\approx \begin{bmatrix} \eta \\ 0 \\ 0 \end{bmatrix} \notag
\end{align}

Under this assumption, the precision-weighted prediction error $\xi'_z$ approximates $2\Gamma_z \lambda_z^2 y'_{h,i}$, and thus signals whether neighbors are instantaneously approaching or moving away from the focal agent, where $\xi'_z < 0$ indicates they are approaching and $\xi'_z > 0$ indicates they are moving away. This in turn determines whether the update to the focal agent's action $\mathbf{v}_i$ is repulsive or attractive, as its sign determines the direction of the social force vector $\Delta\hat{\mathbf{r}}$. 
Although the first order distance $y_{h,i}$ does not directly drive action, it does so indirectly through its effect on inference of $\mu'$. If we consider the case when the sensed distance $y_{h,i}$ drops below the setpoint $\eta$, then one can reason through the cascade of prediction errors that ultimately lead to a repulsive force. As a direct consequence of a drop in $y_{h,i}$ below $\mu$, sensory prediction errors $\xi_{z}$ will become negative, whose minimization will require $\mu$ to move below $\eta$. This process in turn incurs slower-moving (negative) model prediction errors $\xi_{\omega}$, whose minimization drives $\mu$ back to its fixed point of $\eta$, given the dynamic constraint for the beliefs to relax to their fixed point. In order to accomplish this upward movement of $\mu$, either the rate of change of $\mu$ or the sensed distance itself must be positive, i.e., $\dot{\mu} > 0$ or $y_{h,i} > 0$. In the absence of positive $y_{h,i}$, model prediction errors will drive $\mu'$ (and hence $\dot{\mu}$) above $0$. This temporarily sets a larger radius of repulsion, i.e., a larger range of $y'_{h,i}$ for which $\xi'_{z}$ is negative and for which repulsive forces impact the focal agent's velocity. This causes the agent to move away from its neighbors and thus further increase $y_{h,i}$, under the assumption that the agent's prediction of the distance dynamics are correlated with the true change in $x_{h,i}$. Belief updating and action thus work together to accelerate the return of $\mu$ towards $\eta$ and $\tilde{\xi}_{z}, \tilde{\xi}_{\omega}$ towards $0$; for this reason active inference is often described as an account of action and perception driven by `self-fulfilling prophecy' \cite{buckley2017free}.

In order to imbue action with a more direct coupling to the neighbors` distances as is done in the classical self-propelled particle models, rather than the velocity of the distance, one could hand-craft the sensorimotor contingency term $\nabla_{\mathbf{v}_i}\tilde{y}_{h,i}$ to enforce a coupling between $y_{h,i}$ and $\mathbf{v}_i$. This would render the action rule equivalent to a `soft'-form of PD control \cite{baltieri2019pid}, where errors on both the first order state $(y_{h,i} - \mu) \approx (y_{h,i} - \eta)$ and its derivative $(y'_{h,i} - \mu') \approx y'_{h,i}$ would drive changes to the velocity. 

\subsection{Extending to multiple sensory sectors}

The results of the previous sections can be straightforwardly extended to the multivariate case as explored in the main text. The focal agent now senses the local distance computed across a set of distinct sensory sectors. For the model explored in the current work, we split up the computation of the local distance variable into a set of $L$ sensory adjacent sectors that comprise an arc of a given angle, relative to the agent's heading vector $\mathbf{v}_i$.  We define the multivariate distance hidden state as follows (dropping the focal agent index $i$ from the sector-specific hidden states to avoid subscript overload):

\begin{align}
    \mathbf{x}_{h,i} &= \begin{bmatrix} x_{h,1} \\ x_{h,2} \\
    \vdots \\ x_{h,L}
    \end{bmatrix} \\ \notag
    \textrm{where} \hspace{2mm} x_{h,l} &\triangleq \frac{1}{K_l}\sum_{j\in N_l} \| \Delta \mathbf{r}_{ij} \| \label{eq:multivariate_x_GP}
\end{align}

where $N_l$ is the set of neighbors in the $l$\textsuperscript{th} sensory sector, and $K_l = \vert N_l \vert$ (c.f., \eqref{eq:definition_of_x_h}). As with the scalar hidden state defined above, we also equip the vector of sector distances $\mathbf{x}_{h,i}$ with corresponding sector-specific, generalized observations $\tilde{\mathbf{y}}_{h,i}$, i.e.

\begin{align}
    \mathbf{y}_{h,i} &= \mathbf{x}_{h,i} + \mathbf{z}_{h,i} \\ \notag 
    \mathbf{y}'_{h,i} &= \mathbf{x}'_{h,i} + \mathbf{z}'_{h,i} \notag \\
    \textrm{where} \hspace{5mm} p(\tilde{\mathbf{z}}_{h,i}) &= \mathcal N(\tilde{\mathbf{z}}_{h,i}; \mathbf{0}, \tilde{\Sigma}_{\mathbf{z},h})
\end{align}

such that the focal individual now observes a vector of local (noise-perturbed) distances and their first orders of motion. Note that the generalized covariance matrix here $\tilde{\Sigma}_{\mathbf{z}, h}$ is now a $2L \times 2L$ size matrix, that encodes the covariance structure between sector-specific noise and their generalized orders. For all simulations we generated uncorrelated noise across the different sectors, although spatially-smooth noise could be modelled by introducing off diagonal elements in $\tilde{\Sigma}_{\mathbf{z}, h}$, i.e., $\mathbb{E}[z_{h,l} z_{h,k}] \neq 0$.

The agent's generative model is also extended to the multivariate state-space formulation we began with, using a vector of generalised hidden states $\tilde{\mathbf{x}} = (\tilde{x}_1, \tilde{x}_2, ..., \tilde{x}_L)$ to estimate the local distance within each sensory sector. Belief-updating consists in updating a vector of generalised means $\tilde{\boldsymbol{\mu}}$ through integration of \eqref{eq:filtering_equations_specific}.

The action update has an identical form as before, except now the sensorimotor contingency term $\nabla_{\mathbf{v}_i}\tilde{\mathbf{y}}_{h,i}(\mathbf{v_i})$ is a collection of partial derivative vectors, one for each sensory sector:

\begin{align}
    \frac{d\mathbf{v}_i}{dt}
    &= -\nabla_{\mathbf{v}_i}\tilde{\mathbf{y}}_{h,i}(\mathbf{v}_i) ^{\top} \nabla_{\tilde{\mathbf{y}}_{h,i}}F_L  \notag \\
    \nabla_{\mathbf{v}_i}\tilde{\mathbf{y}}_{h,i}(\mathbf{v_i}) &= \begin{bmatrix} \nabla_{\mathbf{v}_i}\mathbf{y}_{h,i}(\mathbf{v}_i)\\\nabla_{\mathbf{v}_i}\mathbf{y}'_{h,i}(\mathbf{v}_i)\end{bmatrix} = \begin{bmatrix} \mathbf{0} \\ \vdots \\ \mathbf{0} \\ \nabla_{\mathbf{r}_i} x_{h,1} \\ \nabla_{\mathbf{r}_i} x_{h,2} \\ \vdots \\ \nabla_{\mathbf{r}_i} x_{h,L}\end{bmatrix}
\label{eq:action_jacobian} 
\end{align}

The last $L$ rows of this Jacobian matrix encode the gradients of the sector-specific distance velocities $y'_{h,l}$ with respect to the focal agent's action; these partial derivatives are vectors pointing from the average position of the neighbors in sector $l$ towards the focal individual. When we combine the Jacobian matrix in \eqref{eq:action_jacobian} with the sensory prediction error term $\tilde{\mathbf{y}}_{h,i}$ (i.e., the free energy gradients $\nabla_{\tilde{\mathbf{y}}_{h,i}}F_L$), we are left with the following update for the velocity:

\begin{align}
    \frac{d\mathbf{v}_i}{dt} &= \boldsymbol{\xi}'_{z} \cdot \Delta \hat{\mathbf{R}} = \begin{bmatrix} \xi'_{z,1} & \xi'_{z,2} & \hdots & \xi'_{z,L} \end{bmatrix} \cdot \begin{bmatrix} \Delta \hat{\mathbf{r}}_1 \\  \Delta \hat{\mathbf{r}}_2 \\ \vdots \\ \Delta \hat{\mathbf{r}}_L\end{bmatrix}\notag \\ &= \sum_{l=1}^{L}\xi'_{z,l} \Delta \hat{\mathbf{r}}_l = 2\Gamma_z \lambda_z^2 \sum_{l=1}^{L}(y'_{h,l} - \mu'_l)\Delta\hat{\mathbf{r}}_l\\
    \textrm{where} \hspace{2mm} \Delta \hat{\mathbf{r}}_l &= \frac{1}{K_l}\sum_{j \in N_{l}} \frac{\Delta \mathbf{r}_{ij}}{||\Delta \mathbf{r}_{ij}||} \notag
\end{align}

The action thus becomes a weighted sum of `sector-vectors' $\Delta\hat{\mathbf{r}}_l$, which are vectors pointing from the focal agent's position $\mathbf{r}_i$ towards the average position of the neighbors in $N_l$. The weights that scale each $\Delta\hat{\mathbf{r}}_l$ are the precision-weighted prediction errors associated with velocity observations emanating from the appropriate sector $\xi'_{z,l} \propto (y'_{h,l} - \mu'_l)$. The fact we can pull the spatiotemporal precision terms $2\Gamma_z\lambda_z^2$ outside the sum over sector-vectors, inherits from a between-sector independence assumption, built into the agent's sensory likelihood model $P(\tilde{\mathbf{y}}|\tilde{\mathbf{x}})$ (see \eqref{eq:spatial_covariance}). If the generative model allowed for between-sector correlations (i.e. $\Sigma^{\mathbf{z}}$ was not diagonal), then the action update would include cross-terms that couple prediction errors from one sector to the sector-vector from another sector.

An active inference agent equipped with such a multivariate representation of the local neighbor-distances thus engages in a sort of `predictive balancing-act', differentially responding more or less to each part of its sensory field in accordance with how much sensations deviate from their posterior expectations $\mu'_l$, where the sign and degree of this deviation is scored by $\xi'_{z,l}$.

\subsection*{Gaussianity of the generative model}
One may question the use of a Gaussian form for the generalized noises $\tilde{\mathbf{z}}, \tilde{\boldsymbol \omega}$ and a linear form for $\tilde{\mathbf{f}}$ and $\tilde{\mathbf{g}}$. These assumptions guarantee a simple generative model whose free energy gradients (with respect to both the state belief $\tilde{\boldsymbol \mu}$ and action $\mathbf{v}$) are linear in the generalized distance observations $\tilde{\mathbf{y}}$. The reason we use a Gaussian form here, is to achieve those simple, linear free energy gradients, i.e., those which align with the classical social forces seen in SPP models (more specifically, the selective attraction-and-repulsion forces first described in \cite{romanczuk2009collective}); in other words, it is exactly an active inference model equipped with this sort of model (and observations with two generalized coordinates, aka position and velocity) that acts in a way equivalent to being driven by vectorial social forces. Note that this also means the Laplace approximation to the posterior over $\tilde{\mathbf{x}}$ is not an approximation at all, but leads to exact inference.

Existing social force models where the forces are linear in the state, can thus be read as special cases of active inference agents equipped with Gaussian generative models -- this is a close cousin to the relationship between linear PID controllers and Gaussian active inference models derived in \cite{baltieri2019pid}. However, if we were to use a different, non-Gaussian of the generative model (either through breaking Gaussianity of the noise model or by introducing nonlinear forms of $\tilde{\mathbf{f}}, \tilde{\mathbf{g}}$), then the updates to the beliefs and actions would no longer be linear functions of the generalized posterior means $\tilde{\boldsymbol \mu}$ and observations $\tilde{\mathbf{y}}$, but could be arbitrary, nonlinear functions of these variables -- and the resulting social forces would have no guaranteed interpretation in terms of attraction and repulsion. Such extensions are an interesting avenue for future work; one interesting possibility would be to explore new extensions of social forces, where the noises $\tilde{\mathbf{z}}, \tilde{\boldsymbol \omega}$ are assumed to belong to the exponential family and that the dynamics model $\tilde{\mathbf{f}}$ is smooth and contains attracting fixed points (i.e., regions of $\mathbf{\tilde{x}}$ where the derivatives of $\tilde{\mathbf{f}}$ vanish). A obvious consequence of such assumptions, is that when the posterior belief $\tilde{\boldsymbol \mu}$ is near these fixed points, i.e., $\tilde{\boldsymbol{\mu}} \approx \arg \min_{\tilde{\mathbf{x}}} \nabla_{\tilde{\mathbf{x}}} \tilde{\mathbf{f}}(\tilde{\mathbf{x}}) \implies \tilde{\boldsymbol \mu}' = 0$, we would still recover attractive and repulsive social forces whose magnitude would be approximately linear in the distance observations $\tilde{\mathbf{y}}$. This is equivalent to making a locally-quadratic (i.e., Laplace) approximation to the free energy landscape around its minima -- this corresponds to just those points in belief-space where the posterior is well-approximated by a Gaussian.

\section{Alignment forces from active inference on angles}
\label{sec:alignment_forces}

In previous sections we have shown how repulsive and attractive forces emerge from active inference models in which the agent entertains a latent representation of the average local distance between itself and its neighbors, and how its heading direction couples to (the derivative of) that variable. In this section we derive alignment-based social forces, like those that appear in the Reynolds, Couzin, and Vicsek models \cite{reynolds1987flocks, couzin2002collective, vicsek1995novel}, as a special case of active inference, where an agent infers the (cosine) angle between its own heading and that of its neighbors, and acts under the prior belief that this angle tends to $0$.

As before, we start with a generative model that represents a generalised latent variable $\tilde{x}_\phi$ that evolves in time with Gaussian additive fluctuations $\tilde{\omega}_{\phi}$. We use the $\phi$ subscript to distinguish this angle-tracking latent variable from the distance-tracking variable of the previous section. We truncate the generalized representation of this state at second order, i.e. $\tilde{x}_{\phi} = \{x_{\phi}, x'_{\phi} \}$, leading to a dynamical equation and corresponding likelihood of the following form:

\begin{align}
    \dot{x}_{\phi} &= -\alpha_{\phi} (x_{\phi} - 1) + \omega_{\phi} \notag \\
    \dot{x}'_{\phi} &= -\alpha_{\phi} x_{\phi}' + \omega'_{\phi} \notag \\
    \implies p(D\tilde{x}_{\phi}|\tilde{x}_{\phi}) &= \mathcal N(D\tilde{x}_{\phi}; \tilde{f}_{\phi}, \tilde{\Sigma}_{\omega_\phi}) \label{eq:angle_dynamics_GM} \\
    \textrm{where} \hspace{2mm} \tilde{f}_{\phi} &= \begin{bmatrix} -\alpha_{\phi}(x_{\phi} - 1) \\ -\alpha_{\phi}x'_{\phi}\end{bmatrix}, \hspace{2mm} \tilde{\Sigma}_{\omega_{\phi}} = \begin{bmatrix} \sigma^2_{\omega_{\phi}} & 0 \\ 0 & \sigma^2_{\omega'_{\phi}} \end{bmatrix}
\end{align}

The observation model describes a mapping from the $0$\textsuperscript{th}-order state to a corresponding observation thereof, perturbed again by Gaussian innovations:

\begin{align}
    y_{\phi} &= x_{\phi} + z_{\phi} \notag \\
    \implies p(y_{\phi}|x_{\phi}) &= \mathcal N(y_{\phi}; x_{\phi}, \sigma^2_{z_\phi}) \label{eq:angle_observation_GM}
\end{align}

Following the same steps as we did previously for the multivariate, distance-tracking generative model, we can write down the Laplace-approximated variational free energy of this model as a quadratic function of the observations and generalized means $\tilde{\mu}_{\phi}$:

\begin{align}
    F_L &\propto \varepsilon_{z_{\phi}}^{\top}\Pi_{z_{\phi}}\varepsilon_{z_{\phi}} + \tilde{\varepsilon}_{\omega_{\phi}}^{\top}\tilde{\Pi}_{\omega_{\phi}}\tilde{\varepsilon}_{\omega_{\phi}} \notag \\
    \textrm{where} \hspace{2mm} \varepsilon_{z_{\phi}} &\triangleq y_{\phi} - \mu_{\phi} \notag \\ \tilde{\varepsilon}_{\omega_{\phi}} &\triangleq D\tilde{\mu}_{\phi} - \tilde{f}_{\phi} \notag
\end{align}

The agent performs a gradient descent on $F_L$ to infer the value of $\tilde{x}_{\phi}$ in light of sensory observations. This inference is encoded by a Gaussian variational posterior with mean $\tilde{\mu}_{\phi}$. As before, we can tune model parameters such that inference is strongly biased by the dynamics model $\tilde{f}_{\phi}$, where the zeroth-order of motion $\mu_{\phi} \approx 1$. The reason we set the set-point at $1$ becomes evident when we consider the generation of sensory data and actions.

Assume that the focal agent with index $i$ observes the local average cosine angle between its own heading vector $\mathbf{v}_i$ and those of its neighbors $\mathbf{v}_j, j \in N_{in}$, where neighbors are once again defined by membership in some interaction zone\footnote{For notational convenience and because it doesn't change the derivations, we omit observation noise on $y_{\phi}$.}:

\begin{align}
    y_{\phi} &= \frac{1}{K}\sum_{j \in N_{in}}\mathbf{v}_{i}^{\top}\mathbf{v}_{j} = \langle \cos(\theta_{ij}\rangle_{N_{in}}
\end{align}

where the equivalence between the dot products and the cosine angle is assured when we assume all $\mathbf{v}_k, k \in \{i\} \cup N_{in}$ have unit magnitude. Recall that if two unit-magnitude vectors $\mathbf{v}_i, \mathbf{v}_j$ are parallel, their dot product (cosine angle) is 1. When we once again assume that agents act by adjusting their heading direction, then the action update given the continuous active inference rule in \eqref{eq:action_update_suppl} has the following form:

\begin{align}
    \frac{d\mathbf{v}_i}{dt} &= -\frac{1}{\sigma^2_{z_{\phi}}}(y_{\phi} - \mu_{\phi}) \hat{\mathbf{v}} \approx (1-y_{\phi}) \hat{\mathbf{v}}  \\
    \textrm{where} \hspace{2mm} \hat{\mathbf{v}} &= \frac{1}{K}\sum_{j \in N_{in}}\mathbf{v}_j
\end{align}

The approximation in the first line holds when we assume the sensory variance $\sigma^2_{z_{\phi}}$ is $1$ and the dynamics prior (either via increasing $\alpha$ or decreasing $\sigma^2_{\omega_\phi}$) dominates inference such that $\mu_{\phi} \approx 1$. In this case, the focal agent $i$ then updates its velocity using the average neighbor velocity. This is proportional to the alignment force in e.g. \cite{vicsek1995novel, couzin2002collective}, except that it is also scaled by how unaligned the focal individual is with its neighbourhood, scored by $1 - y_{\phi}$.

\section{Online parameter estimation}
\label{sec:smoothness_updating}

In this section we derive update rules for the generative model parameters using a simple gradient descent scheme on the Laplace-approximated variational free energy. In the active inference literature this process of updating parameters, as opposed to beliefs about states, is often analogized to online learning or neural plasticity \cite{friston2016active,dacostaActiveInferenceDiscrete2020}.

\subsection{Updating sensory smoothness}
In this section we derive an update equation for the sensory smoothness parameter $\lambda_z$, which captures the generative model's assumptions about the temporal autocorrelation structure of sensory noise $z$.

Recall the formulation of state space models in generalized coordinates of motion in Section \ref{sec:gen_filt}. In addition to providing a concise description of local paths of the state $\vec{x}_t$ in terms of its higher derivatives $x', x'', ..., x^{[n]}$, stochastic differential equations in generalized coordinates also allow one to express \textit{serial correlations} in the noises at the first order $z$, by assuming that it can be differentiated (has non-zero, smooth autocovariance) and represented in terms of hierarchical or generalized noises $z', z'', z''', ..., z^{[n]}$. 

Recall the parameterization of the generalized sensory precision $\tilde{\Pi}^{z}$ as a factorization into two precision matrices, that respectively represent agent's beliefs about the `spatial' and `temporal' covariance structure. We parameterize these with the two precision parameters $\Gamma_z$ and $\lambda_z$. $\Gamma_z$ encodes the agent's belief about the overall magnitude of the fluctuations, and $\lambda_z$ encodes beliefs about their their serial correlations in time, assuming a Gaussian form for their autocorrelation:

\begin{align}
    \tilde{\Pi}^z &= S(\lambda_z) \otimes \Pi(\Gamma_z)
     \notag \\
  \Pi(\Gamma_z) &= \begin{bmatrix}\Gamma_{11} & \\
                             & \Gamma_{22} & & \\
                             & & \ddots & \\
                             & & & \Gamma_{LL}
                            \end{bmatrix} \notag \\
    S(\lambda_z) &= \begin{bmatrix} 1 & 0 & -\frac{1}{2\lambda^2_z}  &\hdots \\
    0 & \frac{1}{2\lambda^2_z} &0 & \\
    -\frac{1}{2\lambda^2_z} & 0 & \frac{3}{4\lambda^4_z} & \\
    \vdots & & & \ddots
    \end{bmatrix}^{-1} \label{eq:generalized_precisions}
\end{align}

We implement a form of behavioral plasticity by allowing agents to update $\lambda_z$ using observations. We accomplish this using a gradient descent on variational free energy:

\begin{align}
    \frac{d\lambda_z}{dt} &= -\kappa_{\theta} \frac{\partial F}{\partial \lambda_z}
\end{align}

where the `learning rate' $\kappa_{\theta}$ is typically set to be at least an order of magnitude lower than the update rate of inference $\kappa_{\mu}$; in all simulations we use $\kappa_{\theta} = 0.001$ and $n_{\text{LearnIter}}=1$ iteration. This enforces a separation of timescales that is typical in generalized filtering and state-space models that perform simultaneous state- and parameter-estimation \cite{friston2008DEMvariational, balaji2011bayesian, baltieri2019pid}.

To compute the gradients of the variational free energy with respect to $\lambda_z$, we can start by expressing those components of the (Laplace-approximated) variational free energy that depend on $\lambda_z$:

\begin{align}
    F(\lambda_z) &= \tilde{\varepsilon}_{\mathbf{z}}^{\top}\tilde{\Pi}^{\mathbf{z}} \tilde{\varepsilon}_{\mathbf{z}} - \ln \left(\det{\tilde{\Pi}^{\mathbf{z}}} \right)
\end{align}

where we only have included the terms that depend on the sensory precision $\tilde{\Pi}^z$ due to its dependence on $\lambda_z$. The full gradient is then simply:

\begin{align}
    \frac{\partial F}{\partial \lambda_z} = \frac{\tilde{\varepsilon}_{\mathbf{z}}^{\top}\tilde{\Pi}^{\mathbf{z}} \tilde{\varepsilon}_{\mathbf{z}}}{\partial \lambda_z} - \frac{\partial \ln \left(\det \tilde{\Pi}^{\mathbf{z}}\right)}{\partial \lambda_z}
\end{align}

Starting with the case of a single sensory sector $L = 1$, then the generalized prediction error $\tilde{\varepsilon}_{\mathbf{z}}$ is a vector of prediction errors, one for each order of motion: $\tilde{\varepsilon}_{z} = \{\varepsilon_{z}, \varepsilon'_{z}, \varepsilon''_{z}, ...\}$ where a sensory prediction error at a given order of motion is simply: $\varepsilon^{[n]}_{z} = y^{[n]} - \tilde{g}^{[n]}$, where the $n$ subscript refers to an order of differentiation. In the case of $3$ generalized coordinates for the simple scalar case:

\begin{align}
    \frac{\partial F}{\partial \lambda_z} &= 4\Gamma_z \lambda_z (\varepsilon'_{z})^2 + \varepsilon''_{z} (8\Gamma_z \varepsilon''_{z} \lambda_z^3 + 2 \Gamma_z \varepsilon_{z} \lambda_z) + 2 \Gamma_z \lambda_z \varepsilon_{z} \varepsilon''_{z} - \frac{6}{ \lambda_z} \notag \\
    &= 4\Gamma_z \lambda_z (\varepsilon_{z} \varepsilon''_{z} + (\varepsilon'_{z})^2 + 2\lambda_z^2 (\varepsilon''_{z})^2) - \frac{6}{\lambda_z}
\end{align}

Meaning that the update for the $\lambda_z$ parameter can be simplified to (omitting the learning rate $\kappa_{\theta}$):

\begin{align}
    \frac{d\lambda_z}{dt} &= -4\Gamma_z \lambda_z(\varepsilon_{z} \varepsilon''_{z} + (\varepsilon'_{z})^2 + 2\lambda_z^2(\varepsilon''_{z})^2) + \frac{6}{\lambda_z}
\end{align}

In the case of the distance-tracking generative model we explore in the main text, we assume that the agents can only observe the $0$\textsuperscript{th} (position, $y$) and 1\textsuperscript{st} (velocity, $y'$) orders of motion of the hidden states $\tilde{x}$. This means there are no longer 2\textsuperscript{nd}-order prediction errors $\varepsilon''_z$ and the update becomes even simpler:

\begin{align}
    \frac{d\lambda_z}{dt} &= -4\Gamma_z\lambda_z(\varepsilon'_{z})^2 + \frac{6}{\lambda_z} \notag \\
     &\approx -4\Gamma_z\lambda_z (y'_{h,i})^2 + \frac{6}{\lambda_z}
\end{align}

where approximation in the second line results in the case of `biased' inference, i.e., $\mu \approx \eta \implies \mu' \approx 0$, allowing us to replace the velocity prediction error $y'_{h,i} - \mu'$ with $y'_{h,i}$.

Given that spatial and temporal precisions are independent from each other due to the factorization of the generalized precision matrix, and further given the diagonal structure of the spatial precision $\Pi_{\mathbf{z}}$ (i.e., independence in random fluctuations across sensory sectors), we can write an update for $\lambda_z$ that is a sum of squared prediction errors across sensory sectors:

\begin{align}
    \frac{d\lambda_z}{dt} &\approx   -4\Gamma_z\lambda_z \sum_{l=1}^{L}(y'_{h,l})^2 - \frac{6L}{\lambda_z}
\end{align}

The quadratic form of this update means that the update to the smoothness parameter decays in proportion with the overall magnitude of the velocity prediction errors, regardless of its sign. This means that if the distance is fluctuating quickly in any direction, then the agent will infer that fluctuations are slightly-less serially-correlated at the $0$\textsuperscript{th} order, reflected by a decrease in $\lambda_z$.

\section{Adding a target representation into the generative model}
\label{sec:target_generative_model}

As described in the main text, it is straightforward to add an additional observation model and dynamics model to an agent's generative model to represent the distance between itself and some abstract spatial target, which in the context of the collective information transfer experiments, we represent with \textbf{T}:

\begin{align}
    \dot{x}_{\text{target}} &= -\alpha_{\text{target}}x_{\text{target}} + \omega_{\text{target}} & y_{\text{target}} &= x_{\text{target}} + z_{\text{target}} \notag\\
    \dot{x}'_{\text{target}} &= -\alpha_{\text{target}}x'_{\text{target}} + \omega'_{\text{target}} & y'_{\text{target}} &= x'_{\text{target}} + z'_{\text{target}} 
\end{align}

We truncate the generalized hidden states at third order $\tilde{x}_{\text{target}} = (x_{\text{target}},x'_{\text{target}}, x''_{\text{target}})$ and the observations at second order $\tilde{y}_{\text{target}} = (y_{\text{target}},y'_{\text{target}})$. When the agent assumes the generalized noises $\tilde{\omega}_{\text{target}}$ and $\tilde{z}_{\text{target}}$ are zero-mean and normally-distributed with covariances $\tilde{\Sigma}^{\omega_{\text{target}}}$ and  $\tilde{\Sigma}^{z_{\text{target}}}$ and leverage the Laplace approximation exactly as we did in the previous section, then we can supplement the Laplace-approximated free energy in \eqref{eq:laplace_vfe_specific} with additional terms corresponding to target-related prediction errors:

\begin{align}
    F_{L} &\propto \frac{1}{2}\left[\tilde{\boldsymbol{\varepsilon}}_{z\text{-Soc}}^{\top}\tilde{\Pi}^{\boldsymbol{z}\text{-Soc}}\tilde{\boldsymbol{\varepsilon}}_{z\text{-Soc}} + \tilde{\boldsymbol{\varepsilon}}_{\omega\text{-Soc}}^{\top}\tilde{\Pi}^{\boldsymbol{\omega}\text{-Soc}}\tilde{\boldsymbol{\varepsilon}}_{\omega\text{-Soc}} + \tilde{\varepsilon}_{z\text{-Tar}}^{\top}\tilde{\Pi}^{z\text{-Tar}}\tilde{\varepsilon}_{z\text{-Tar}} + \tilde{\varepsilon}_{\omega\text{-Tar}}^{\top}\tilde{\Pi}^{\omega\text{-Tar}}\tilde{\varepsilon}_{\omega\text{-Tar}} \right] + C
\end{align}

Here we use the suffixes "$\text{-Soc}$" or "$\text{-Tar}$" to indicate `social' relevant information (related to the average neighbor distance) and the `target' prediction errors. $C$ captures all the additional terms (log determinants of precision matrices, etc.) that are constant with respect to the posterior means $\tilde{\boldsymbol{\mu}} = (\tilde{\boldsymbol{\mu}}_{\text{Social}}, \tilde{\mu}_{\text{Target}})$. Following the same reasoning as used to derive the inference and action rules for the case of the social distance hidden states and observations($\tilde{\mathbf{x}}_{\text{Social}}, \tilde{\mathbf{y}}_{\text{Social}})$, we can do the same to derive active inference rules for the target-relevant hidden states and observations $\tilde{x}_{\text{target}}, \tilde{y}_{\text{target}}$:

\begin{align}
\frac{d \tilde{\boldsymbol{\mu}}_{\text{Social}}}{d t} &= D\tilde{\boldsymbol{\mu}}_{\text{Social}} - \nabla_{\tilde{\boldsymbol{\mu}}_{\text{Social}}}F_L(\tilde{\boldsymbol{\mu}}_{\text{Social}}, \tilde{\mathbf{y}}_{\text{Social}})&
\frac{d\mathbf{v}}{dt} &= -\left(\nabla_{\mathbf{v}}F_L(\tilde{\boldsymbol{\mu}}_{\text{Social}}, \tilde{\mathbf{y}}_{\text{Social}}) + \nabla_{\mathbf{v}}F_L(\tilde{\mu}_{\text{Target}}, \tilde{y}_{\text{Target}})\right)\notag \\
\frac{d \tilde{\mu}_{\text{Target}}}{d t} &= D\tilde{\mu}_{\text{Target}} - \nabla_{\tilde{\mu}_{\text{Target}}}F_L(\tilde{\mu}_{\text{Target}}, \tilde{y}_{\text{Target}}) 
\label{eq:active_inference_equations_target}
\end{align}

Where expanding the free energy gradients on the right equation leads to an expression for the action update in terms of a precision-weighted sum of vectors, appearing in \eqref{eq:social_target_weighted_expression} in the main text.



\section{Numerical methods}
\label{sec:numerical_solutions}

We used a forwards Euler-Maruyama scheme to the integrate a (Itô-style) stochastic differential equation for the positions of all agents over time:

\begin{align}
    d\mathbf{r}_t &= \mathbf{v}_t dt + \sigma_a dW_t
\end{align}

where the variance of `action noise' $\sigma^2_a$ was set to $0.01$ for all experiments unless explicitly stated otherwise. We used a step size of $\Delta t = 0.01 s$ in the integration. For the current timestep $\tau$ in `simulation time', we used a simple forwards Euler scheme to integrate the differential equations used for belief updating (see \eqref{eq:filtering_equations_specific}) and action (see \eqref{eq:velocity_update_active_inference}) for each agent in parallel. We use the positions and heading vectors of all agents from the previous integration timestep ($\tau - \Delta t$) to generate the observations for the current timestep.

The collective information transfer experiments were performed using custom Julia code, and all other simulations were implemented in JAX using custom code. To accelerate the parameter scans over $p_{inf}$, $\Gamma_{z\text{-Social}}$, and $\Gamma_{z\text{-Target}}$ to create the results in Figure \ref{fig:collective_acc_as_pinformed_gammazsoc_heatmap_accur_both_precisions} in the main text, we used the high-performance computing clusters (Cobra and Draco) provided by the Max Planck Computing and Data Facility.

\begin{table}\centering
\caption{Default parameter configuration used in numerical simulations (unless otherwise stated). First column denotes the name of the parameter, second column denotes its default value and third column indicates whether the parameter concerns the generative process (the physics of the simulation environment), the generative model used for active inference, or a hyperparameter (e.g., used in the active inference algorithm).}

\begin{tabular}{lrrr}
Parameter & Value & Type \\
\midrule
$\Delta t$ (Euler integration step, in seconds) & $0.01$ & generative process \\
Number of sensory sectors & $4$& generative process\\
Sector angle (in degrees $^\circ$) & $60$ & generative process \\
$R_{0}$ (interaction radius, in arbitrary units) & $5$ & generative process\\
$\sigma^2_{a}$ & $0.01$ & generative process\\
$\sigma^2_{z,h}$& $0.01$ & generative process\\
$\sigma^2_{z',h}$ & $0.01$& generative process \\
$\Gamma_z$ & $1.0$ & generative model\\
$\Gamma_{\omega}$ & $1.0$ & generative model\\
$\lambda_z$ & $1.0$ & generative model\\
$\lambda_{\omega}$ & $1.0$ & generative model\\
$\alpha$ & $0.5$& generative model\\
$\alpha_{\text{target}}$ & $0.5$& generative model\\
$\eta$ & $1.0$ & generative model\\
Number of generalised coordinates ($x$) & $3$ & hyperparameter \\
Number of generalised coordinates ($y$) & $2$ & hyperparameter \\
$\kappa_{\mu}$ & $0.1$ & hyperparameter\\
$n_{\text{InferIter}}$ & $1$ & hyperparameter\\
$\kappa_{a}$ & $0.1$ & hyperparameter\\
$n_{\text{ActionIter}}$ & $1$ & hyperparameter\\
$\kappa_{\theta}$ & $0.001$ & hyperparameter\\
$n_{\text{LearnIter}}$ & $1$ & hyperparameter\\
\bottomrule
\end{tabular}
\label{table:tableD1}
\end{table}

\begin{sloppypar}
\printbibliography[title={Supplemental References}]

\begin{figure*}
\centering
\includegraphics[width=1.0\linewidth]{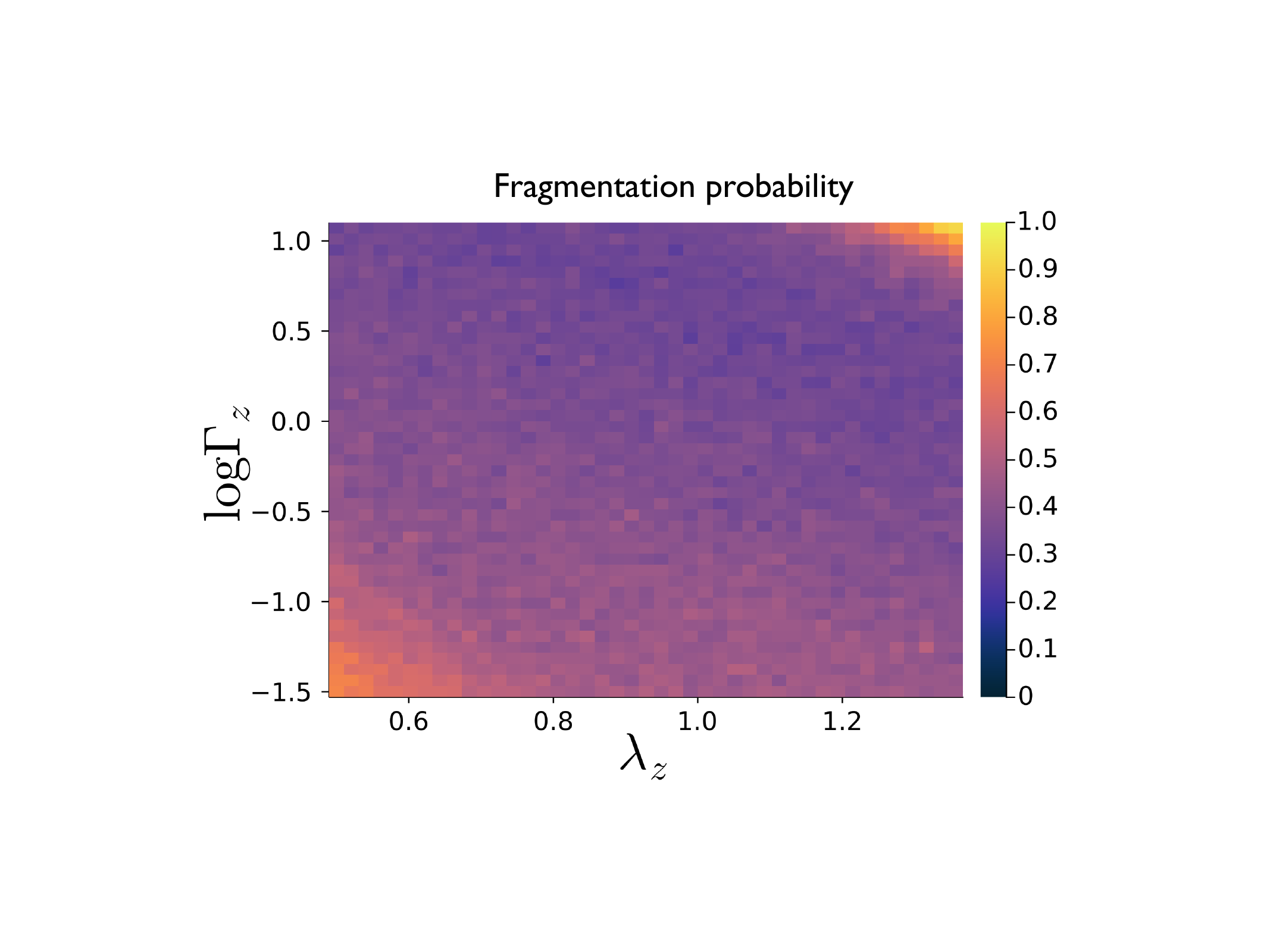}
\caption{Fragmentation probability as a function of the two precision parameters $\log\Gamma_{z}$ and $\lambda_z$. Fragmentation probability was quantified as the proportion of trials (out of 500 independent trials per condition) where the group fragmented. A trial was considered fragmented if least one individual was further than 2.0 dimensionless units from all other individuals for at least 3 of the last 10 seconds of the 15-second trial. All other parameters are identical to those used in Figure \ref{fig:collective_regimes_collective_properties_analysis}B in the main text.}
\label{fig:fragmentation_probability}
\end{figure*}

\section*{Supplemental Movie Legends}
\paragraph*{Movie 1} Example of a simulation of $N=96$ agents that includes a dynamic transition from polarized to milling regime. Parameters: $\sigma^2_{z',h} = 0.05$; Sector angle $=80^\circ$; $R_{0}=10$ units; $\kappa_{a} = 0.2$; $\lambda_{\omega} = 0.5$; $\lambda_z = 2.0$. Unless specified, all remaining parameters are as listed in Table \ref{table:tableD1}.


\paragraph*{Movie 2} Example of a polarized group of  $N=64$ agents. Parameters: Sector angle $=80^\circ$;;  $\kappa_{a} = 0.2$; $\lambda_{\omega} = 0.5$; $\lambda_z = 1.5$.


\paragraph*{Movie 3}Example of a milling regime observed in $N=64$ agents. Parameters: $\sigma^2_{z',h} = 0.04$;  Sector angle $=80^\circ$; $\alpha = 1.0$; $\kappa_{a} = 0.2$; $\lambda_{\omega} = 0.8$; $\lambda_z = 1.8$.


\paragraph*{Movie 4} Example of a disordered regime observed in $N=96$ agents. Parameters: Number of sensory sectors $=2$; Sector angle $=160^\circ$; $R_{0}=10.0$ units; $\alpha = 0.2$; $\eta = 0.5$, $\kappa_{a} = 0.2$; $\lambda_{\omega} = 0.1$; $\lambda_z = 1.787$.


\paragraph*{Movie 5} Metastable `snaking' configuration observed in $N=64$ agents. Parameters: $\sigma^2_{z',h} = 0.04$; Sector angle $=80^\circ$; $\alpha = 0.1$; $\kappa_{a} = 0.2$; $\lambda_{\omega} = 0.5$; $\lambda_z = 2.2$.


\end{sloppypar}
\end{refsection}

\end{document}